\documentclass{article}
\usepackage[a4paper,left=2.54cm,right=2.54cm,top=2.54cm,bottom=2.54cm]{geometry}
\usepackage[utf8]{inputenc}
\usepackage{xspace}
\usepackage{amsfonts}
\usepackage{comment} 
\usepackage{amsmath,bm}
\usepackage[nolist]{acronym}
\usepackage{graphicx}
\usepackage{xspace}
\usepackage{siunitx}
\usepackage{hyperref}
\usepackage{lineno}

\newcommand{\mymath}[2]{\newcommand{#1}{\TextOrMath{$#2$\xspace}{#2}}}
\mymath{\frcLossWeight}{\lambda_\mathrm{FRC}}
\mymath{\ssimLossWeight}{\lambda_\mathrm{SSIM}}
\mymath{\gpLossWeight}{\lambda_\mathrm{GP}}

\mymath{\encoder}{\mathrm{\mathbf E}}
\mymath{\mlp}{\mathrm{\mathbf F}}
\mymath{\discriminator}{\mathrm{\mathbf D}}
\mymath{\generator}{\mathrm{\mathbf G}}
\mymath{\dataDistribution}{p_D}
\mymath{\randomDistribution}{p_\nu}
\mymath{\latentVector}{\mathbf{z}}
\mymath{\firstAngle}{\alpha}
\mymath{\secondAngle}{\beta}
\mymath{\viewAngle}{\nu}
\mymath{\realAngle}{v}
\mymath{\predictContrast}{\mathbf{\hat{c}_\nu}}
\mymath{\realContrast}{\mathbf{c}_v}
\mymath{\contrast}{\mathbf{c}}
\mymath{\spatialCoordinates}{\textit{\textbf{x}}}
\mymath{\temporalCoordinates}{\textit{t}}
\mymath{\refractiveIndex}{n}

\begin{acronym}
\acro{DL}{Deep Learning}
\acro{IoR}{Index of Refraction}
\acro{ONIX}{Optimized Neural Implicit X-ray imaging}
\acro{NeRF}{Neural Radiance Fields}
\acro{CNN}{Convolutional Neural Network}
\acro{MLP}{Multilayer Perceptron}
\acro{MSE}{Mean Squared Error}
\acro{GAN}{Generative Adversarial Network}
\acro{XMPI}{X-ray Multi-Projection Imaging}
\acro{SASE}{Self-Amplified Spontaneous Emission}
\acro{ESRF}{European Synchrotron Radiation Facility}
\acro{SR}{Synchrotron Radiation}
\acro{MHz-XMPI}{MHz X-ray Multi-Projection Imaging}
\acro{PSF}{Point Spread Function}
\acro{MAE}{Mean Absolute Error}
\acro{ML}{Machine Learning}
\acro{FRC}{Fourier Ring Correlation}
\acro{FSC}{Fourier Shell Correlation}
\acro{SSIM}{Structural Similarity Index Measure}
\acro{3D}{Three Dimensional}
\acro{XFEL}{X-ray Free-Electron Laser}
\acro{European XFEL}{European X-Ray Free-Electron Laser Facility}
\acro{SPring-8}{Super Photon Ring – 8 GeV}
\acro{ONIX}{Optimized Neural Implicit X-ray imaging}
\acro{CNN}{Convolutional Neural Network}
\acro{MLP}{Multilayer Perceptron}
\acro{GAN}{Generative Adversarial Network}
\acro{DSSIM}{Dissimilarity Structure Similarity Index Metric}
\acro{PDE}{Partial Differential Equation}
\acro{CDI}{Coherent Diffraction Imaging}
\acro{PINN}{Physics-Informed Neural Network}
\acro{IoR}{Index of Refraction}
\end{acronym}

\def\figurePath{Figures/}

\newcommand{\refFig}[1]{Figure~\ref{fig:#1}}
\newcommand{\refEq}[1]{Equation~\ref{eq:#1}}
\newcommand{\name}{4D-ONIX\xspace}

\def\figurePath{Figures/}

\def\myfigure#1#2{%
    \begin{figure}[htbp!]%
    \centering\includegraphics*[width = 0.8\linewidth]{\figurePath#1}%
    \vspace{-.2cm}%
    \caption{#2}%
    \label{fig:#1}%
    \end{figure}%
}

\def\mycfigure#1#2{%
    \begin{figure*}[htbp!]%
    \centering\includegraphics*[width = \linewidth]{\figurePath#1}%
    \vspace{-.2cm}%
    \caption{#2}%
    \label{fig:#1}%
    \end{figure*}%
}

\newcommand{\change}[2]{%
\hypertarget{hyperline:#1}{%
\def\temp{#1}\ifx\temp\empty%
\else%
\linelabel{line:#1}%
\fi%
\textcolor{black}{#2}}%
}

\newcommand{\test}[1]{%
  \hfil\penalty0 \hfilneg
  \textcolor{black}{\texttt{#1}}%
}
\newcommand{\namestar}{4D-ONIX$^*$\xspace}
\newcommand{\ntoi}{Noise2inverse\xspace}

\title{\name: A deep learning approach for reconstructing 3D movies from sparse X-ray projections}
\author{Yuhe Zhang$^{1}$, Zisheng Yao$^1$, Robert Klöfkorn$^2$,Tobias Ritschel$^3$, and Pablo Villanueva-Perez$^1$}
\date{%
    $^1$Synchrotron Radiation Research and NanoLund, Lund University, Box 118, 221 00, Lund, Sweden\\%
    $^2$Center for Mathematical Sciences, Lund University, Box 117, 221 00, Lund, Sweden\\%
    $^3$University College London, WC1E 6BT London, UK\\[2ex]%
}

\begin{document}

\maketitle

\begin{abstract}
The X-ray flux provided by X-ray free-electron lasers and storage rings offers new spatiotemporal possibilities to study in-situ and operando dynamics, even using single pulses of such facilities. 
\ac{XMPI} is a novel technique that enables volumetric information using single pulses of such facilities and avoids centrifugal forces induced by state-of-the-art time-resolved 3D methods such as time-resolved tomography.
As a result, \ac{XMPI} \change{}{offers the potential to }acquire 3D movies (4D) at least three orders of magnitude faster than current methods.
However, it is exceptionally challenging to reconstruct 4D from highly sparse projections as acquired by \ac{XMPI} with current algorithms.
Here, we present \name, a  \ac{DL}-based approach that learns to reconstruct 3D movies from an extremely limited number of projections. 
It combines the computational physical model of X-ray interaction with matter and state-of-the-art \ac{DL} methods.
We demonstrate the potential of \name to generate high-quality 4D by generalizing over multiple experiments with only two \change{}{to three projections per timestamp for binary droplet collisions and additive manufacturing.}
We envision that \name will become an enabling tool for 4D analysis, offering new spatiotemporal resolutions to study processes not possible before.
\end{abstract}

\section*{Introduction}
X-ray tomography is a non-destructive 3D imaging technique that enables the study of the internal structure and composition of opaque materials~\cite{stock1999x, maire2014quantitative,withers2021x}.
It has been widely applied in various research areas, such as materials science~\cite{baruchel2000x,ebner2013visualization}, medical imaging~\cite{hendee1999physics}, biology~\cite{mizutani2012x}, geology~\cite{mees2003applications}, fluid dynamics~\cite{bieberle2016combined,schug2016imaging}, and also industrial diagnostics~\cite{ostman1988application,hussain2017review}.
By utilizing X-rays to scan a sample from multiple angles, tomography generates a comprehensive 3D image of the object.
The recent advancement of modern \ac{SR} facilities and \acp{XFEL}  has opened the way for time-resolved tomography experiments, where exploring dynamics in 4D, i.e., in real-time and in 3D space, has become possible~\cite{villanova2017fast}.
Time-resolved tomography experiments have demonstrated the capabilities of capturing dynamics with sub-millisecond temporal resolution and micrometer spatial resolution~\cite{shahani2020characterization, garcia2021tomoscopy}.
However, the imaging process of tomography introduces centrifugal forces to the sample during acquisition that can potentially alter or even damage the sample and the dynamics studied. 
Achieving \SI{1}{\milli\second} temporal resolution would necessitate rotating the sample 500 times per second, generating a substantial centrifugal force that is hundreds of times the gravitational acceleration. 
This rapid rotation also presents a challenge in developing sample environments that can withstand such speeds.
Consequently, the necessary rotation process for collecting a full set of projections often restricts the types of samples that can be used and limits the temporal resolution of X-ray tomography experiments. 
Additionally, the nature of tomography is not adaptable to single-shot imaging approaches.
Various techniques have been developed to overcome this limitation, which aim to achieve higher temporal resolutions while preserving high-quality 3D images of the sample.
\acf{XMPI}~\cite{hoshino2011development, hoshino2013development, villanueva2018hard, voegeli2020multibeam,patent} is developed among others as a time-resolved imaging technique.
Unlike conventional scanning-based techniques, \ac{XMPI} records volumetric information without scanning by generating different beamlets that illuminate the sample simultaneously from different angles.
By combining the concept of \ac{XMPI} with the unique capabilities of the fourth generation  \ac{SR} sources~\cite{bilderback2005review,shin2021new} and \acp{XFEL}~\cite{mcneil2010x}, one can record 3D information of dynamical processes from kHz~\cite{liang2023sub,asimakopoulou2023development} up to MHz rate~\cite{villanueva2023megahertz}, exploiting the possibility of imaging 3D using single-pulses of \acp{XFEL}.
This opens up new possibilities for studying the high-speed dynamics of various materials. 
However, unlike tomography which records hundreds of projections of a sample, \ac{XMPI} records no more than eight projections due to practical constraints~\cite{hoshino2011development,villanueva2018hard,asimakopoulou2023development}, and a reconstruction algorithm is required to recover 4D (3D + time) information from the extremely sparse data collected from \ac{XMPI}.

It is unlikely to solve this problem using traditional methods due to the highly ill-defined nature of the problem~\cite{jacobsen2019x}.
\change{}{Classic approaches to this problem typically rely on matching low-level primitives or features from different projections~\cite{bellucci2023hard,duarte2019computed,fainozzi2023three,ippoliti2022reconstruction}. However, these methods require easily identifiable features or prior knowledge and assumptions about the sample, which limits their applicability. In addition, their performance degrades when applied to complex objects.}
Recent advancements in \ac{DL} approaches provide a potential solution to this problem.
\ac{DL} algorithms, such as \acp{CNN}~\cite{lecun2015deep} and \acp{GAN}~\cite{goodfellow2020generative}, can be \change{}{optimized} to learn the underlying structure of the sample\change{}{, generalizing over different similar samples,} and produce high-quality reconstructions from sparse inputs~\cite{ulyanov2018deep, ledig2017photo,ahishakiye2021survey}. 
Specifically, approaches based on \ac{NeRF}~\cite{Mildenhall2020NERF} have recently shown promise in optical and X-ray imaging for reconstructing high-resolution 3D / 4D structures from sparse views~\cite{schwarz2020graf,yu2021pixelnerf, Pumarola_2021_CVPR, Yuhe2022ONIX,maas2023nerf,corona2022mednerf,zheng2023ultrasparse}.
Instead of relying on voxels, these methods learn the shape of an object as an implicit function of the 3D spatial coordinates, offering a potential solution to the longstanding memory issues associated with 3D reconstructions.
The recently developed ONIX algorithm showed the 3D reconstruction from eight views for the experimental and simulated tomographic experiments~\cite{Yuhe2022ONIX}.
However, there is a need for a 4D reconstruction algorithm to investigate ultrafast dynamical processes using highly sparse X-ray projections acquired through \ac{XMPI}.

Here, we report \name, a self-supervised \ac{DL} model that learns to reconstruct high-quality temporal and spatial information of the sample from the extremely sparse projections collected with \ac{XMPI}. 
It does not require 3D ground truth or prior dynamic description of the sample at any stage - neither during the training nor during its deployment.
Once trained, it can reconstruct a 3D movie showing the refractive index of the sample as a function of time from only the recorded projections.
The capability of \name is achieved by i) incorporating the physics of X-ray propagation and interaction into the model, ii) having a continuous representation of the sample that describes the refractive index as a function of position and time, iii) learning the latent features of the sample by generalizing over all timestamps, and iv) applying adversarial learning to enforce consistency between measured and predicted projections.
We demonstrate our approach on dynamical processes of binary water droplet collisions~\cite{adam1968collision,planchette2012onset,grzybowski2014modelling} \change{}{and additive manufacturing~\cite{makowska2023operando}.
First, we validate the performance of our approach using simulated droplet collision datasets modeled using the Navier-Stokes Cahn-Hilliard equations~\cite{Houseini:17, Lovric2019LowOF}, retrieving volumetric information from two projections of simulated \ac{XMPI} experiments.
We then validate the approach on experimental data of additive manufacturing, reconstructing the melt pool dynamics of magnetite-modified alumina from three projections.}
This approach has also been applied to experimental \ac{XMPI} data collected at \ac{ESRF} and \ac{European XFEL} with kHz up to MHz acquisition rates,  ~\change{}{with results reported in other studies} ~\cite{asimakopoulou2023development, villanueva2023megahertz}.
We envision that \name will be pivotal for the implementation and applications of \ac{XMPI}, and it will enable new spatiotemporal resolutions for time-resolved 3D X-ray imaging through novel acquisition approaches based on sparse projections.
The 4D reconstructions offered by \name will provide valuable observations for in-situ and operando testing for a plethora of systems, e.g., the characteristics and dynamic studies in fluid dynamics and material science, which are important for various applications such as the study of atmospheric aerosols~\cite{wang2008ultrafast,loh2012fractal}, advancements in fuel cell technologies~\cite{buhrer2020unveiling} and improvements in additive manufacturing~\cite{martin2019dynamics,martin2019ultrafast,kumar2023machine,jin2020machine}. 
It is worth mentioning that our approach can also be extended to single-shot phase contrast imaging~\cite{olbinado2017mhz,hagemann2021single,zhang2021phasegan} and coherent diffraction imaging~\cite{rodriguez2015three,fan2023coherent} experiments where the propagation model is explicitly known.
Furthermore, the availability of 4D reconstruction from \name opens up the possibility of directly constraining the dynamics in the reconstruction process, e.g., through the use of \acp{PINN}~\cite{RAISSI2019686}.

\section*{Results}
\mycfigure{main}{
Demonstration of \ac{XMPI} experiment~\cite{patent,villanueva2023megahertz} and the reconstruction approach.
\textbf{a} Conceptual illustration of the \ac{XMPI} setup. The dashed blue box on the right shows the goal of the reconstruction approach.
\textbf{b} Overview of the \name approach.
}

\subsection*{Concept of \ac{XMPI} and overview of the approach}

The concept of \ac{XMPI} is depicted in \refFig{main} \textbf{a}.
Unlike tomography measurements which rotate the sample in a period of 180\textdegree ~or 360\textdegree ~and record, typically, hundreds to thousands of projections in between, the measurement of \ac{XMPI} does not require rotations of the sample.
\ac{XMPI} relies on high-brilliance X-ray sources and a group of beam splitters to generate multiple beams that illuminate the sample simultaneously from different angles and a set of kHz/MHz detectors to record the different sample projections. 
In this way, one can record volumetric information on the fast dynamics of the sample, ultimately limited by the speed of the detector and the flux of the X-ray source.
Inset marked by the blue box of \refFig{main} \textbf{a} shows the goal of \ac{XMPI}, which is to achieve a continuous 3D movie of the sample being studied from the projections recorded, exploiting excellent temporal characteristics provided by \acp{XFEL} or the fourth-generation  \ac{SR} sources.
This opens up possibilities for observing in 3D kHz up to MHz dynamics of the sample.
As shown in \refFig{main} \textbf{a}, the data used here comes from applying XMPI to water droplet collisions using two split beamlets.

We designed a self-supervised \ac{DL} algorithm, \name, to reconstruct temporal and spatial information from \ac{XMPI}.
It combines neural implicit representation~\cite{Mildenhall2020NERF} and generative adversarial mechanism~\cite{goodfellow2020generative} with the physics of X-ray interaction with matter, resulting in a mapping between the spatial-temporal coordinates and the distribution of the refractive index of the sample. 
By enforcing consistency between the recorded projections and the estimated projections generated by the model, the model learns by itself the 3D volumetric information of the sample at each measured time point from only the given projections without needing real 3D information about the sample.
Once trained, it provides a 3D movie showing the structure and dynamics of the sample.

\subsection*{Self-supervised 4D reconstruction approach}
The architecture of the self-supervised 4D reconstruction model (\name) is depicted in \refFig{main} \textbf{b}.
The goal of the model is to learn the \ac{IoR} of the sample at any spatial-temporal point.
The \name model is based on three neural networks: an encoder, a 4D IoR (index of refraction) generator, and a discriminator.
The encoder and the discriminator are built up on \acp{CNN}, whereas the IoR generator is formed by fully connected \acp{MLP}.
The IoR generator learns the local features of each object, while the encoder and the discriminator capture both the local features and global features across all timestamps and experiments.
\change{}{
The recorded projections are first passed through the encoder, which converts the 2D images into stacks of downscaled feature maps.
For each spatial-temporal point, we apply an affine coordinate transformation to transfer from the global coordinate system to the local coordinate system of each camera. 
This transformation allows us to retrieve the corresponding features extracted by the encoder.
}
The IoR generator receives extracted feature maps from the encoder and predicts the value of the index of refraction at each spatial point.
In this way, we can predict the value of the index of refraction of the sample at any spatial-temporal point.
Please note that the IoR generator relies on neural implicit representation, which learns a continuous function of the refractive index.
It differs from many common 3D reconstruction approaches that learn the value of each isolated voxel.
We apply a physics-based forward propagation model to predict projections from different angles in the plane of incoming X-rays. 
The forward model is based on the projection approximation (weak scattering)~\cite{Paganin2006CoherentX-ray}, where secondary scattering caused by the X-ray photons is ignored.
Note that this can also be extended to multiple scattering conditions by using multi-slice methods~\cite{Paganin2006CoherentX-ray}.
The discriminator constrains the predictions by differentiating them from the real ones and minimizing the differences between them.
The networks are trained by an adversarial loss function.
We use the same encoder, generator, and discriminator for the training of the sequence of all 3D movies.
Therefore, the networks are trained by all of the recorded projections so that these networks, especially the convolutional layers in the encoder and the discriminator, learn generalized features across the whole sequence.
Further details of the network implementation and the loss function are included in the \nameref{subsec:algorithm} and \nameref{subsec:training details} sections.

\begin{figure}[htb]%
\centering\includegraphics*[width =0.5 \linewidth]{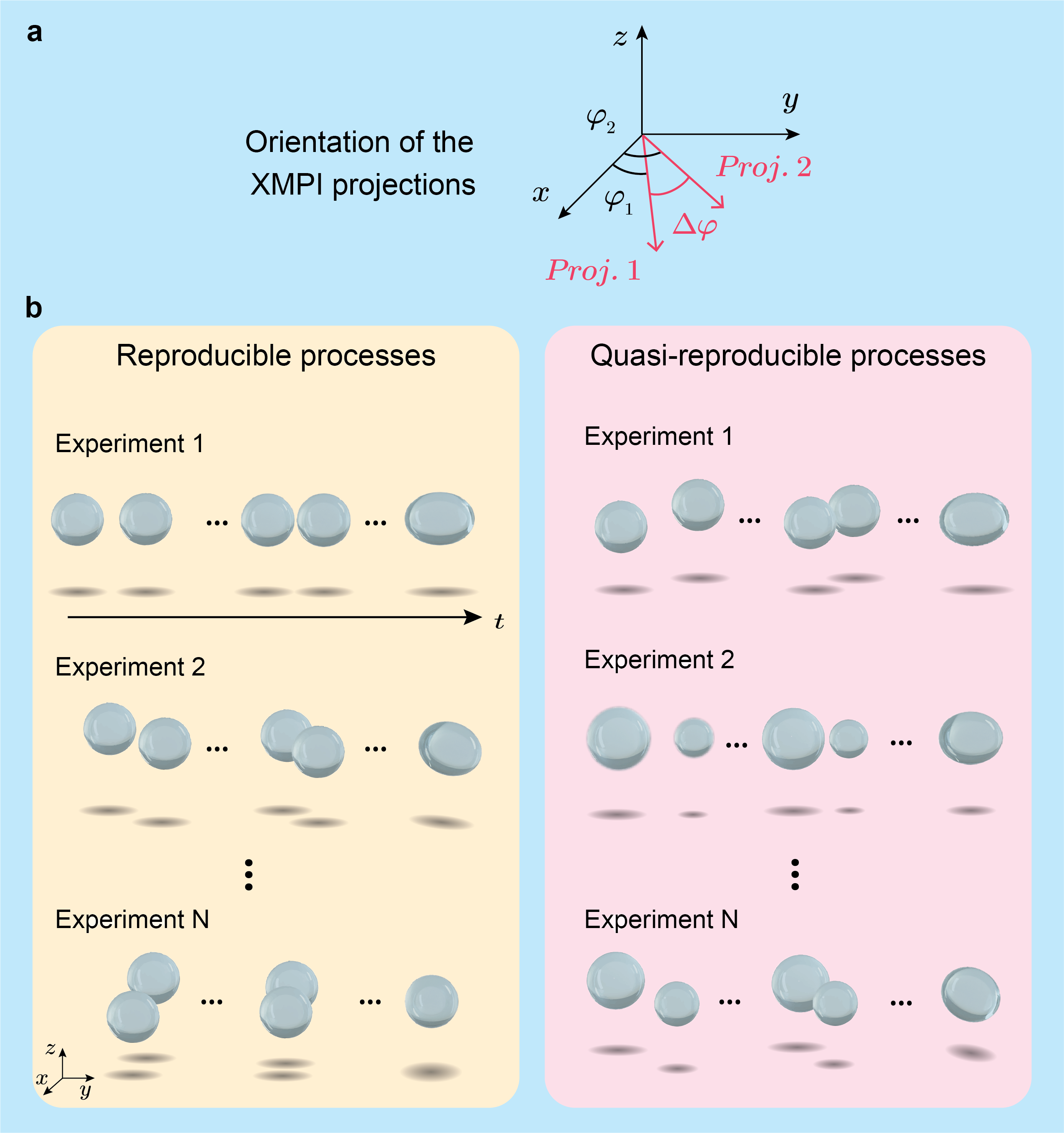}%
\vspace{-.2cm}%
\caption{Demonstration of the simulated \ac{XMPI} data. \textbf{a} Geometry of the simulated \ac{XMPI} projections. Projection pairs are generated on the x-y plane with a fixed angle between them. \textbf{b} Comparison of the two training scenarios. In the reproducible processes scenario, the processes are identical, while in the quasi-reproducible processes scenario, dynamical processes can differ. For example, the size of droplets may vary among experiments, as shown here.}%
\label{fig:scenarios}%
\end{figure}

\change{}{\subsection*{\name demonstration on simulated water droplet collisions}}
\label{subsec:simulation_results}
We assess the performance of \name using simulated datasets of droplet collisions modeled with the Navier-Stokes Cahn-Hilliard equations.
We refer the readers to the \nameref{subsec:simulation method} section for details about the droplet collision simulation.
The simulations provided multiple sequences of 3D objects, each sequence illustrating the collision process between two water droplets.
Two projections were generated from each 3D object to mimic the experimental conditions of the XMPI experiment performed at the \ac{European XFEL}~\cite{villanueva2023megahertz}.
For brevity, we refer to each collision sequence as an XMPI experiment or simply an experiment.
The geometry of the projection generation in the simulated XMPI experiment is illustrated in \refFig{scenarios} \textbf{a}.
We evaluate \name using the simulated datasets under two scenarios, as visualized in \refFig{scenarios} \textbf{b}. 

\begin{enumerate}
    \item Reproducible processes. It involves multiple identical experiments of the same dynamical process; as shown in \refFig{scenarios} \textbf{b}, the samples may exhibit a variety of orientations throughout the experiments. 
    This can arise from various factors such as random sample orientations, samples arriving from different directions, or manual rotation of the sample stage.
    For instance, assuming we set the projection from detector 1 of the first experiment as \ang{0} and denote the relative angle between the two projections as $\Delta \varphi$, in the first experiment, the dynamics process is measured at $\varphi_1=\ang{0}$ and $\varphi_2=\Delta \varphi$.
    In the second experiment, the sample orientation is shifted by $\phi = \ang{30}$, and the two projections are measured at $\varphi_1^\prime=\ang{30}$ and $\varphi_2^\prime=\ang{30}+\Delta \varphi$. 
    The third experiment can be measured at $\varphi_1^{\prime\prime}=\ang{60}$ and $\varphi_2^{\prime\prime}=\ang{60}+\Delta \varphi$, and so on.
    This allows for obtaining volumetric information on the collision process from different angles of the sample without rotation.
    \item Quasi-reproducible processes. In many cases, measuring perfectly reproducible processes is challenging or impractical. It is more realistic to measure several experiments capturing similar dynamical processes within experimental tolerances, with each process being measured only once, i.e., resulting in only one 4D sequence available for each process.
    As illustrated in \refFig{scenarios} \textbf{b}, similar to the first scenario, the dynamical processes are measured from different orientations.
    However, the dynamical processes themselves are not identical.
    For example, in the first experiment, both droplets have the same size, whereas in the second experiment, one droplet is larger than the other. In the third experiment, the droplets may be moving faster than in the first experiment, and so forth. 
\end{enumerate}

\mycfigure{simulation}{Demonstration of the 4D reconstruction for a reproducible process. Seven timestamps are shown, illustrating different stages of the droplet collision process. The two generated projections are shown in rows 1-2, corresponding to \ang{0} and \ang{23.8}, respectively. The top view and side view of the 3D ground truth are shown in rows 3-4. The top and side views of the \name reconstructions are shown in rows 5-6. The bottom two rows show the MSE and DSSIM between the \name reconstruction and the ground truth for each corresponding timestamp.}

Two training datasets were generated for the two scenarios.
The first dataset was based on a single simulation, mimicking 16 experiments of a reproducible process. Each experiment contains a pair of projections with random $\varphi$ angles and fixed $\Delta \varphi$ for each timestamp.
The second dataset was simulated for quasi-reproducible processes. 
It was based on 16 simulations, where a single projection pair with random orientation of the sample was selected from each simulation to form a training dataset. 
All 16 simulations simulated the collision process of binary droplets,  with 10\% variance in collision velocities and 10\% variance in droplet size based on the tolerance of the droplets collision experiment conducted at European XFEL~\cite{villanueva2023megahertz}.
For both datasets, the relative angle between the projection pairs was selected as \ang{23.8} to match the experimental conditions of the XMPI setup at the European XFEL~\cite{villanueva2023megahertz}.
Each simulation contained 75 timestamps, and both datasets contained 1200 timestamps in total.

We trained \name with different numbers experiments, from 1, 2, 4, 8, to 16, for both scenarios.
The reconstruction results and their quality improvement as a function of increasing the number of experiments are presented in the Supplementary Information.
Here, we only present the best results trained with 16 experiments.
First, we present the results of training with reproducible processes.
\refFig{simulation} shows examples of the \name performance trained on reproducible processes.
It shows the two projections, the 3D of the ground truth and the reconstruction rendered from the side and top views.
The side view corresponds to the x-axis, while the top view aligns with the z-axis, which is perpendicular to the acquisition plane.
Seven timestamps are shown, demonstrating the crucial stages in the collision process for one of the collision experiments. 
In this collision experiment, two identical water droplets were accelerated and collided center-to-center at the same speed.  
A full movie of the reconstruction is provided in Supplementary Information.
Please note that the 3D ground truth shown here was only used for comparison, and they were never shown to the network at any stage.
The reconstructions were quantified by \ac{MSE} and \ac{DSSIM}~\cite{Wang2004SSIM}, and their values for selected timestamps are shown in \refFig{simulation}. 
Both of the metrics calculate the difference between the \name reconstructions and the ground truth. 
Therefore, a value closer to zero indicates a better reconstruction.  
Furthermore, we analyzed the evolution of errors throughout the collision process for each 3D timestamp, with the results displayed in Figure S1 (Supplementary Information).
In addition to 3D metrics, we also computed metrics in the 4D domain.
We calculated the 4D MSE and 4D DSSIM for all 75 timestamps of the demonstrating experiment, resulting in MSE=$2.6\times10^{-4}$ and DSSIM=$2.3\times10^{-3}$.
Apart from evaluating data correlation in the object space, we also evaluated the performance of the reconstructions in the frequency space using \ac{FSC} and \ac{FRC}~\cite{saxton1982correlation,VANHEEL2005250}. 
\ac{FSC} and \ac{FRC} calculate the normalized cross-correlation between the reconstructions and the ground truth in frequency space over shells (3D) and rings (2D), respectively. 
The resolution of the reconstructions can be retrieved from the correlation curves.
Here, we used the half-bit threshold criterion for resolution determination.
The retrieved 3D spatial resolution was 4 $\pm$ 1 voxels,  with the voxel side size being equivalent to the pixel size of the input projections.
The 2D resolution was separately calculated for the training views and unseen views. For the unseen views, resolution was 6.0 $\pm$ 1.6 pixels, while for the training views, it was 4.7 $\pm$ 1.6 pixels.
Please refer to Figure S2 (Supplementary Information) for the plots of \ac{FSC} and \ac{FRC} for example timestamps.

\mycfigure{simulation_multiple_exp}{Demonstration of the 4D reconstruction for a quasi-reproducible processes. In this demonstrating experiment, the droplet on the right is bigger and moves faster than the one on the left. 
The seven timestamps shown here depict different stages of the droplet collision process.
The two projections are shown in rows 1-2, corresponding to \ang{0} and \ang{23.8}, respectively. The top view and side view of the 3D ground truth are shown in rows 3-4. The top and side views of the \name reconstructions are shown in rows 5-6. The bottom two rows show the MSE and DSSIM between the \name reconstruction and the ground truth for each corresponding timestamp.}

Second, we present the results obtained from training with the dataset of quasi-reproducible processes, comprising multiple simulations featuring similar but not identical samples.
\refFig{simulation_multiple_exp} demonstrate the performance of \name trained on quasi-reproducible processes.
Analogously to the reproducibility test, the figure presents the two projections, the 3D representation of the ground truth, and the reconstruction rendered from side and top views, providing a representation of the collision process between two water droplets.
A full 3D movie of the reconstruction can be found in the Supplementary Information.
In this specific experiment, one of the water droplets (the one on the right) was 7\% larger than the other, and the larger droplet also moved 7\% faster than its counterpart.
The 4D metrics for the reconstructions were calculated, resulting in  MSE=$4.3\times10^{-4}$ and DSSIM=$3.2\times10^{-3}$. 
The distribution of errors throughout the collision process for each timestamp is visualized in Figure S1 (Supplementary Information).
We determined the 3D spatial resolution using \ac{FSC}, resulting in a resolution of 6 $\pm$ 1 voxels. 
The 2D resolution was found to be 7 $\pm$ 2 pixels for the training views and 10 $\pm$ 4 pixels for the unseen views.
Please refer to Figure S4 (Supplementary Information) for additional figures of \ac{FSC} and \ac{FRC}.

\test{ 
\subsection*{\name demonstration on experimental additive manufacturing data}}
\label{subsec:exp result}
\change{}{
We also validated the performance of our reconstruction method using experimental data. 
As current XMPI experiments have only been applied to unknown dynamical processes and achieve spatiotemporal resolutions beyond what existing methods can offer, ground truth data for evaluation is unavailable. 
Instead, we validated our approach using experimental data from time-resolved X-ray tomography, or tomoscopy~\cite{garcia2021tomoscopy}.
The tomoscopy experiments were conducted at the TOMCAT beamline of the Swiss Light Source~\cite{makowska2023operando}, capturing the melt pool dynamics of magnetite-modified alumina.
The projections were recorded every \ang{0.9} during continuous sample rotation at 50 Hz, and in total 200 projections were recorded within \ang{180} for each tomogram~\cite{aluminadata2023}.}
\change{}{
We simulated a reproducible \ac{XMPI} experiment for the remelting of magnetite-modified alumina using 60 tomograms that resulted in 60 time points. 
For each experiment and time point, we selected three projections spaced \ang{27} apart.
A total of 32 \ac{XMPI} experiments were simulated with three projections each. 
For example, the projection angles for the first experiment were \ang{0}, \ang{27}, and \ang{54}, while for another experiment they were \ang{3.6}, \ang{30.6}, and \ang{57.6}, and so on.
Please note that these experiments were treated as independent in the context of \name. 
As a result, \name only sees the three projections spaced \ang{27} apart and does not have access to the relative angles between experiments as the latter is, in general, not experimentally possible. 
Each projection had an effective pixel size of \SI{2.75}{\micro\meter} and a field of view of 912 $\times$ 180. 
We selected a $960 \times 64$ region where the remelting dynamics occur and resized this area to $128 \times 64$ for faster computation.
For more details on the experimental geometry and data preparation, please refer to the Supplementary Information.
}

\change{}{
\refFig{exp_result_psi} demonstrate the performance of \name on the experimental additive manufacturing data.
It presents the projection triplets from an example experiment, along with the 3D representation of the ground truth and the reconstruction, rendered from both side and top views, and the performance metrics for the selected timestamps.
As before, the top view was perpendicular to the projection plane and was never shown to the networks.
Four time points for the remelting process of magnetite-modified alumina are shown. 
The dynamic remelting regions are highlighted with blue boxes in both the top and side views, while red circles indicate areas that pose challenges for the algorithm.
A full 3D movie of the reconstruction and the ground truth can be found in the Supplementary Information.
The 4D metrics for the reconstructions were calculated, resulting in MSE=$5.2\times10^{-3}$ and DSSIM=$6.9\times10^{-2}$. 
For the distribution of 3D MSE and DSSIM over time, as well as the performance of \name trained with varying numbers of experiments, please refer to Table S3 and Figure S7 in the Supplementary Information.
We also calculated the 3D spatial resolution using \ac{FSC}, yielding a resolution of 2 voxels in the resized 128 $\times$ 64 spaces over all time points. 
}

\change{}{
We also conducted additional evaluations by comparing \name to two well-established sparse-view tomographic reconstruction methods: SART~\cite{andersen1984simultaneous} (Simultaneous Algebraic Reconstruction Technique), a classic method for sparse-view tomographic reconstruction, and Noise2inverse~\cite{hendriksen2020noise2inverse}, a state-of-the-art deep learning-based approach.
Since these methods lack the capacity to generalize across different experiments, we reconstruct using a hypothetical scenario where more projections are available for a single experiment.
The results of \name trained with more projections of a single experiment are also compared, as denoted by \namestar.
The results are presented in Figure S8 and S9 in the Supplementary Information.
}

\mycfigure{exp_result_psi}{Demonstration of \name on experimental data.
Four timestamps are shown, illustrating different stages of the additive manufacturing process. 
The three projections are shown in rows 1-3, corresponding to \ang{0}, \ang{27}, and \ang{54}, respectively. 
Note that the scales of the horizontal and vertical directions differ, as we resized the image for faster computation and improved visualization.
The top view and side view of the 3D ground truth are shown in rows 4-5. 
The top and side views of the \name reconstructions are shown in rows 6-7. 
The blue boxes mark the remelting regions in both the top and side views, while the red circles indicate an example area that poses challenges for the reconstruction algorithm. 
The bottom two rows show the MSE and DSSIM between the \name reconstruction and the ground truth for each corresponding timestamp.
}

\section*{Discussion} 
We have demonstrated the application of \name, a self-supervised \ac{DL} 4D reconstruction model to reconstruct 4D data of water droplet collisions using simulated \ac{XMPI} datasets \change{}{and additive manufacturing using experimental datasets.}

\change{}{For the simulated water droplet collision data,} we evaluated the performance of \name under two scenarios: with reproducible processes and with quasi-reproducible processes.
As depicted in \refFig{simulation} and  \refFig{simulation_multiple_exp}, we reconstructed 3D movies of water droplet collision using 16 experiments.
Each experiment contained only two 2D projections over 75 different timestamps. 
The reconstructions accurately captured the dynamics of the water droplet collision process.
Even the top view, which was perpendicular to the projection plane, was effectively reconstructed.
This view was experimentally unobservable and never shown to the networks.
We evaluated the \ac{MSE}, \ac{DSSIM}, and the resolution of the reconstructions.
The MSE was $2.6\times10^{-4}$ for the training with reproducible processes and $4.3\times10^{-4}$ for the training with quasi-reproducible processes.
The DSSIM was $2.3\times10^{-3}$ and $3.2\times10^{-3}$ for the training with reproducible and quasi-reproducible processes, respectively.
This suggests that our approach effectively reconstructs the collision processes of the water droplets. 
The ability to reconstruct 3D from two projections is largely attributed to \name's capacity to generalize across all timestamps of various experiments using an encoder. 
Furthermore, \name learns not only the self-consistency of the two projections in 3D but also the shared features of the samples through the discriminator. 
Regarding the resolution of the 3D reconstructions, \name achieved a resolution of approximately 4 voxels for training with reproducible processes and 6 voxels for training with quasi-reproducible processes.
Further investigation found that the reconstructions had a superior resolution for the trained views over unseen views for both scenarios.
This is mainly due to the fact that the projections were only recorded in the plane formed by the two beamlets, providing very limited information from the top.
As shown in \refFig{simulation}, \refFig{simulation_multiple_exp}, and Figure S1, the error distribution with time suggested that the timestamps prior to the collision process appear to be more difficult to reconstruct. 
This difficulty may come from the distortion of the droplets induced by the acceleration process within the simulation and the inadequacy of training data available for this particular stage.
Comparing the training results between reproducible and quasi-reproducible processes, it is observed that the training shows better performance for reproducible processes than quasi-reproducible processes.
Learning the features of a single experiment is a less complex task than generalizing across multiple experiments with different parameters within the experimental tolerance.
Introducing multiple experiments may introduce additional variability and complexity, making it more challenging for the networks to capture and reconstruct essential features.
Nevertheless, obtaining reproducible samples or processes is not always guaranteed, making quasi-reproducible processes a more common scenario and closer to experimental conditions.
We also conducted comparisons by training \name with different numbers of experiments, and the corresponding metrics for the reconstructions are detailed in Table S1 and Table S2.
The results indicated that the performance was less favorable with 1-2 experiments, but improved greatly when more than 4 experiments were available. 
Reconstructions with fewer than four experiments may lead to a range of diverse and incorrect solutions for the sparse-view reconstruction problem.
This variability, particularly noticeable in the top view (the unseen view), resulted in reconstructions that deviated from the ground truth, as illustrated in Figure S3 and Figure S5.
Our results with water droplet collision demonstrated that 16 experiments were adequate for training an accurate model. 

\change{}{
For the experimental additive manufacturing remelting data, we evaluated the MSE, DSSIM, and the resolution of the reconstructions determined by FSC.
The MSE was $5.2\times10^{-3}$, the DSSIM was $6.9\times10^{-2}$, and the spatial resolution was 2 voxels, indicating that our approach effectively captures the remelting processes.
The dynamic remelting regions (marked by blue boxes in \refFig{exp_result_psi}) were well reconstructed, as shown in the upper part of the side views. 
However, some static features were not fully reconstructed.
A notable example is the bottom-right tip of the material, marked by red circles. 
Other imperfections include the lack of flatness in some areas of the reconstruction and an incidental shape generated at the top of the top view. 
These issues may arise due to the complexity and noise of the projections.
The differences are particularly noticeable in the top view, which represents an unseen perspective of the algorithm.
Figure S8 presents a comparison of \name trained with different numbers of experiments, as well as results from SART, Noise2inverse, and \namestar trained with different numbers of projections of a single experiment.
This comparison demonstrates that \name, when trained with 32 experiments, achieves performance levels comparable to \namestar, SART, and \ntoi when they are trained on single experiments with 24 projections.
However, given that the current XMPI setup does not support acquiring 24 projections simultaneously, the ability of \name to generalize across multiple experiments with fewer projections is crucial. 
This generalization capability makes \name particularly suitable for practical experimental conditions, where obtaining a high number of projections may not be feasible.
Additionally, the reconstruction quality differs across methods due to the noisy nature of the experimental projections.
The SART and \ntoi reconstructions are particularly affected by strong artifacts, which can obscure finer details and reduce the accuracy of the reconstruction. 
By contrast, \name and \namestar employ self-consistency constraints across spatial and temporal dimensions, effectively reducing artifacts and enhancing the clarity of reconstructed features. 
This enables \name and \namestar to better mitigate noise, generating reconstructions that are more robust and accurate under challenging experimental conditions. 
As a result, both models show potential for handling complex, noisy datasets in 4D more effectively than SART and \ntoi, offering a promising approach for 4D reconstructions in high-noise environments.
}

\name has also been applied to experimental data collected at \ac{European XFEL}, where collisions of binary droplets were recorded at 10 keV in 1.128 MHz frame rate using \ac{XMPI} with two split beamlets~\cite{villanueva2023megahertz}.
\name can reconstruct a complete 3D movie capturing the collision of water droplets from the experimental \ac{XMPI} dataset, using just two projections for each dynamical process.
The temporal resolution of the 3D movie retrieved was \SI{0.89}{\micro\second}, which is three orders of magnitude faster than state-of-the-art time-resolved tomography~\cite{garcia2021tomoscopy}.
Unfortunately, the retrieved reconstructions may suffer from imperfections as i) only two sequences of the droplet collision were captured, with a minimal shift in sample orientation ($23.8^\circ$), and ii) the recorded projections suffered from noise and imaging artifacts.

\name is a data-driven approach, and its performance is determined by the quality and quantity of the training dataset. 
In this work, \name has been trained using just two \change{}{to three} projections, which is an extremely ill-posed problem that can limit the accuracy of the reconstruction. 
Setups of \ac{XMPI} have also been demonstrated on \ac{SR} sources at \ac{SPring-8} in Japan and \ac{ESRF} in France, which generate more projections and can capture dynamical processes at sub-millisecond timescales with $\sim$ \SI{10}{\micro\metre} spatial resolution~\cite{liang2023sub, asimakopoulou2023development}.
These extra projections, as well as increasing the angular separation between projections ($\Delta\varphi$), can improve the performance of \name and reduce the number of experiments required to satisfactorily reconstruct 4D information.
Thus, such advancements in \ac{XMPI} will improve \name's performance. 
The number of experiments is a crucial factor that affects the accuracy of the reconstruction.
We have evaluated its effect for this specific scientific case, as illustrated in Figures S1, S3, S5 \change{}{and S7}.
One can observe that the model's efficacy may be compromised when trained with only one or two experiments.
Increasing the number of experiments for identical or similar samples is key to improving both convergence and the accuracy of the reconstructions.
This way, the neural networks receive more information about the sample, contributing to improved model performance.

The 4D nature and the inclusion of X-ray physics in \name offer opportunities for further development.
First, the temporal information available from \name's reconstructions provides opportunities for improvement by introducing additional constraints.
For instance, if the sample dynamics adhere to a \ac{PDE}, incorporating an extra \ac{PDE} loss term into the loss function using \acp{PINN}~\cite{RAISSI2019686} can not only better align the model with the laws of physics but also facilitate the interpolation between the dynamical processes.
This will enable the generation of a continuous 3D movie with temporal resolution surpassing the recording rate of the \ac{XMPI} experiment.
Furthermore, the inclusion of a frame variation regularizer can further improve the reconstruction quality~\cite{Pumarola_2021_CVPR}. 
The \name code provides the option to include a frame variation regularizer in the time domain, minimizing the variance of adjacent timestamps. 
This can be particularly beneficial when dealing with high temporal resolutions or rapidly evolving samples.
Second, \name offers adaptability and flexibility to be applied to other types of time-resolved imaging experiments, such as coherent diffraction imaging~\cite{rodriguez2015three,fan2023coherent} and phase-contrast experiments~\cite{olbinado2017mhz,hagemann2021single,zhang2021phasegan}. 
The propagation and interaction model implemented in \name can be readily adapted to suit different experiments' specific imaging formation processes, e.g., to directly perform 4D phase reconstructions.

\change{}{The proposed approach also faces several challenges.
First, our algorithm requires multiple experiments with similar dynamics to achieve high-quality 4D reconstructions. As mentioned in our discussion, the performance of the model improves remarkably when trained on data from several experiments compared to a single experiment. The optimal number of experiments is influenced by several factors, including the number of projections and the angle between them, the total number of timestamps, and the complexity of the process being studied. As a result, the required number of experiments must be determined on a case-by-case basis. Second, our approach assumes varying sample orientations across different experiments to maximize coverage of the Fourier space, as demonstrated in Table S1. The effectiveness of the method may decline if altering the sample orientation between experiments is not feasible, which could limit the reconstruction quality. Third, our model is computationally heavy due to the integration of multiple CNNs, making it challenging for real-time or online reconstructions. While this complexity is necessary for achieving the desired resolution and accuracy, it does present challenges in scenarios where rapid processing is required.
}

In conclusion, we have presented \name, a novel physics-based self-supervised deep learning approach capable of reconstructing high-quality temporal and spatial information of the sample from extremely sparse projections.
\name provides an innovative solution for single-shot time-resolved X-ray imaging techniques like \ac{XMPI}, which rely on recording a sparse number of sample projections simultaneously to avoid the scanning processes, as used in X-ray tomography.
With the application of \name on simulated XMPI data of water droplet collision \change{}{and experimental data of additive manufacturing}, we have demonstrated the capacity of our approach to reconstruct high-resolution 3D movies from only two \change{}{to three} projections per timestamp, preserving the critical dynamics of the observed phenomenon.
We envision that \name will open up unprecedented possibilities for more sophisticated time-resolved X-ray imaging experiments and push the limits of time-resolved imaging when combining the use with \ac{XMPI} and advanced high-brilliant X-ray sources such as the fourth-generation  \ac{SR} sources and \acp{XFEL}.
The reconstructions provided by \name will allow for an in-depth exploration of rapid physical phenomena in fluid dynamics and material science, potentially impacting fields like atmospheric aerosol research and additive manufacturing.
Additionally, the self-supervised learning approach demonstrated by \name has the potential to provide new spatiotemporal resolutions through novel acquisition approaches that only acquire a sparse number of projections.
Finally, the possibility to directly reconstruct 3D processes provides a framework to implement physics-based methods for dynamic studies.

\section*{Methods}

\subsection*{Simulation of water droplet collision}
\label{subsec:simulation method}

\newcommand{\vecu}{\mathbf{u}}

Consider the domain $\Omega$ and time $t \in (0,T]$, denote $\Omega_T = \Omega \times (0,T]$. In this paper, we use $\psi \in [-1,1]$ as the phase variable, with $\psi =1$ to label phase $1$ (i.e. water) and $\psi = -1$ to label phase $2$ (i.e. air), and $\psi \in (-1,1)$ representing the interface.

Consider the non-dimensional Reynolds and Weber numbers defined as 
\begin{equation}
Re := \frac{\rho_r U L}{\mu_r} \quad \text{ and } \quad We := \frac{\rho_r U^2 L}{\sigma}
\quad,
\end{equation}
with $U$ the characteristic velocity, $L$ the characteristic length scale, $\sigma$ the surface tension and $\rho_r = \rho_1$ and $\mu_r = \mu_1$ as the reference density and viscosity for non-dimensionalization.

Following \cite{Houseini:17, Lovric2019LowOF}, 
using volume averaged densities and viscosities 
\begin{equation}
\rho(\psi) = \frac{1}{2} \big ( (1 + \psi)\rho_1 + (1-\psi) \rho_2 \big ) \quad \text{ and } \quad
\mu(\psi) = \frac{1}{2} \big ( (1+\psi)\mu_1 + (1-\psi)\mu_2\big )
\end{equation}
the fluid flow is described by the incompressible Navier-Stokes equations in non-dimensionalized form with potential surface tension $\eta\nabla\psi$
\begin{alignat}{3}
\label{eq:ns1}
\rho(\psi) \big (\partial_t \vecu + \vecu \cdot \nabla \vecu\big) - \frac{\mu(\psi)}{Re} \nabla \cdot \nabla \vecu  + \nabla p &= -\frac{1
}{We} \eta \nabla\psi &\quad \mbox{ in }  \Omega_T,\\
\nabla \cdot \mathbf{u} &= 0  &\quad \mbox{ in }  \Omega_T,
\label{eq:ns2}
\end{alignat}
with an appropriate combination of boundary conditions for velocity $\vecu$ and pressure $p$.

The movement and shape of the interface is described by the Cahn-Hilliard equations, given the double well potential
$W(\psi) = \frac{1}{4}(\psi^2-1)^2$ and thus $\partial_{\psi}W(\psi) = (\psi^2-1)\psi$:

\begin{align}
\label{eq:ch1}
\partial_t \psi + \nabla \cdot (\mathbf{u}\psi - \omega \nabla \eta) &= 0 \quad \mbox{ in }  \Omega_T \\
\eta - \partial_{\psi}W(\psi) - \varepsilon^2 \Delta \psi &= 0 \quad \mbox{ in }  \Omega_T 
\label{eq:ch2}
\end{align}
where $\psi$ is the phase variable, $\eta$ the chemical potential, and $\omega$ a mobility parameter and $\varepsilon$ is an artificial parameter describing the interface thickness. The equations are, in our case, equipped with natural boundary conditions. 

Similar to \cite{Lovric2019LowOF}, low-order Finite Element techniques are used to discretize the Navier-Stokes Cahn-Hilliard equations. For the Navier-Stokes equations \eqref{eq:ns1},\eqref{eq:ns2} a pressure-correction projection method with second order backward differencing formula (BDF2) in time, described in \cite{guermond:00}, and a Taylor-Hood conforming Finite Element pair $(\mathbb{P}_2, \mathbb{P}_1)$ for $(\vecu,p)$ in space are used. 
For the Cahn-Hilliard equation \eqref{eq:ch1},\eqref{eq:ch2} a conforming Finite Element pair $(\mathbb{P}_1,\mathbb{P}_1)$ for $(\psi, \eta)$ and Crank-Nicholson in time with mass lumping is used. The equations are decoupled by solving sequentially first the interface motion using the velocity from the previous time step followed by a Navier-Stokes solve using the updated phase variable. This works well for reasonably small time steps. The resulting linear systems are solved with a (S)SOR preconditioned GMRes or CG method from the DUNE-ISTL module \cite{blatt_bastian_ISTLParallel:2008} with a tolerance of $10^{-8}$. To reduce the computational effort, mesh adaptation based on the newest vertex bisection is applied \cite{dnvb:16, alugrid:16}. The refinement indicator is based on the phase variable $\psi$ and defined as $\theta(\psi) = 2 (\psi + 1) ( 1 - \psi )$.
A mesh element is refined when $\theta(\psi) > 0.15$ and coarsened whenever $\theta(\psi) < 0.0525$.
The implementation is done in the open-source framework DUNE and, in particular, DUNE-FEM \cite{dunereview:21, dunefemdg:21} using the Python based user interfaces, which allows the description of weak forms using UFL \cite{ufl:14}. 

   \begin{figure}[!ht]
   \begin{center}
      \includegraphics[width=0.3\textwidth]{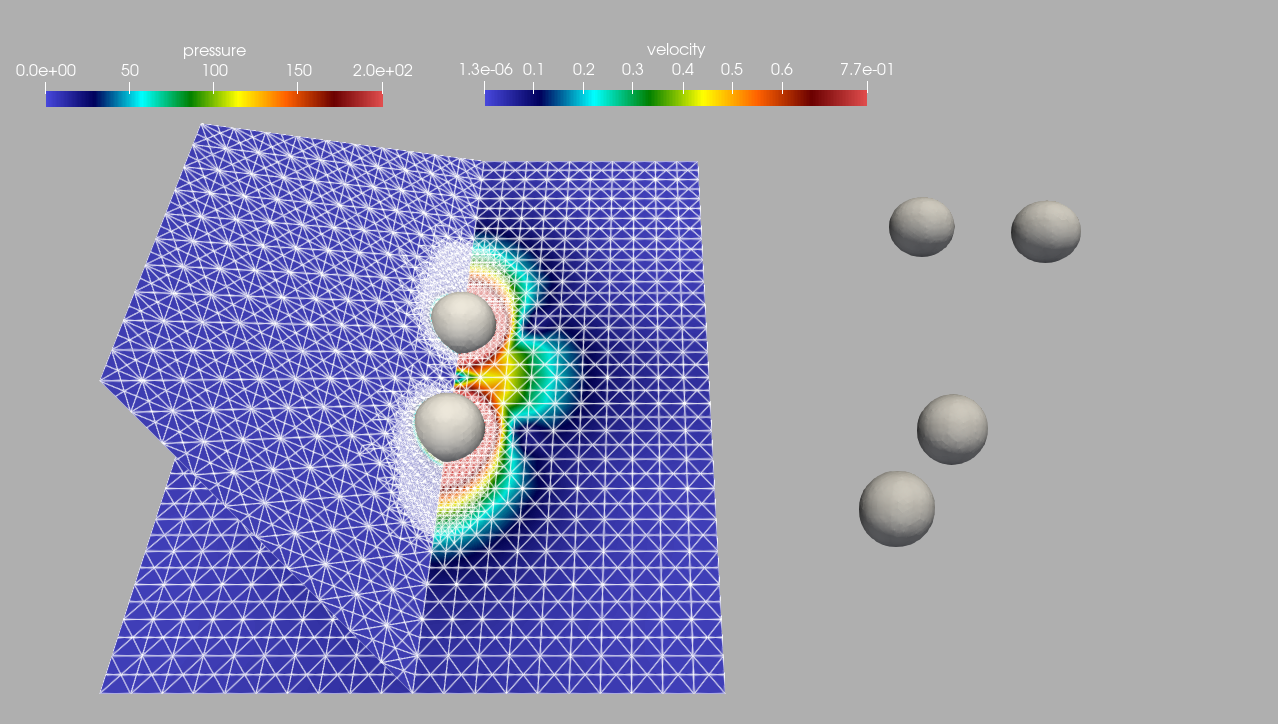}
      \includegraphics[width=0.3\textwidth]{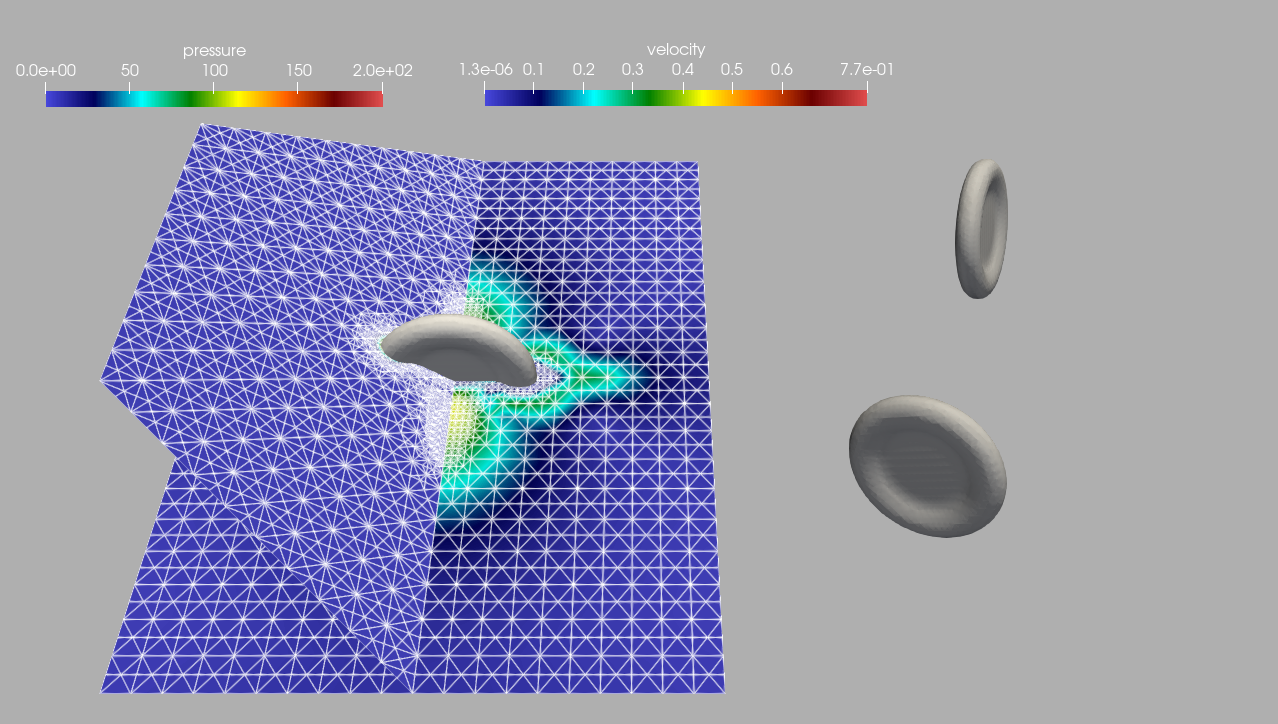}
      \includegraphics[width=0.3\textwidth]{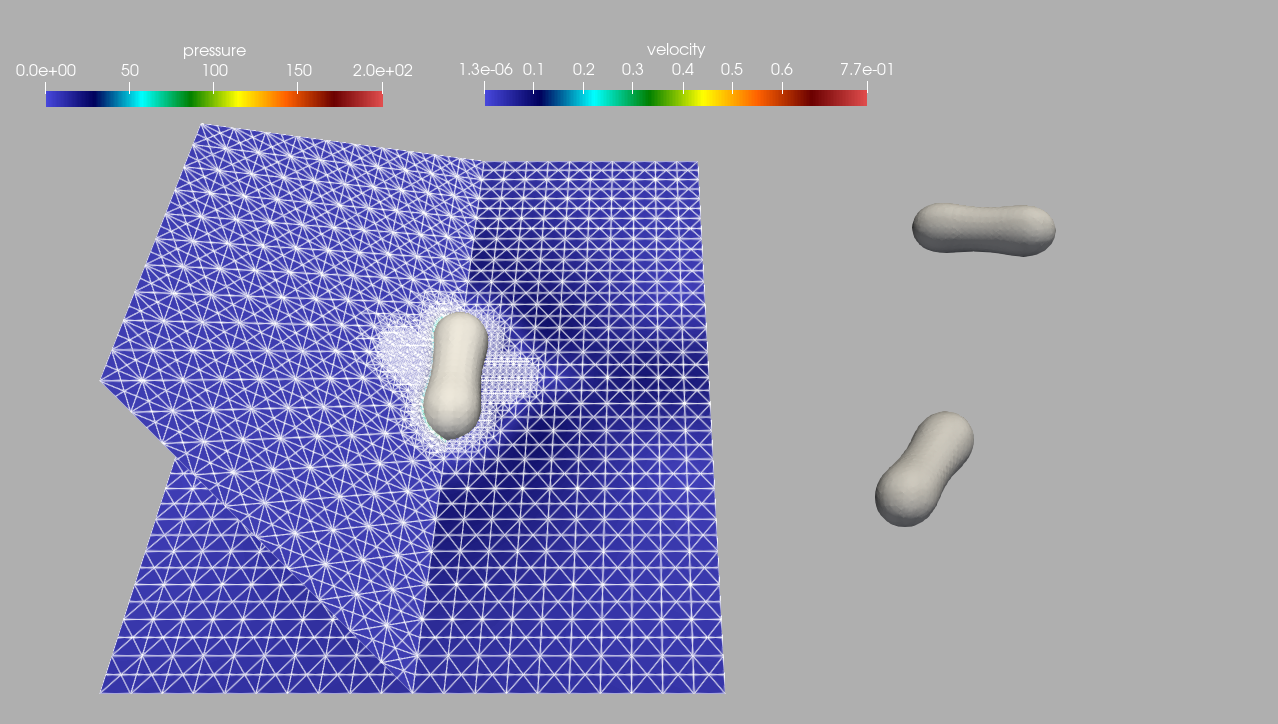}
      \caption{Equal sized centered collision of two droplets viewed from different
      angles. Dynamic grid adaptation is used which reduces the computational
      effort by a factor of $40$.}
     \label{fig:droplet}
   \end{center}
  \end{figure}

For the simulations carried out in this work, parameters resembling the experiment were used. In particular, $Re = 200$, $We = 6.94$, 
$U = 2.5\, m/s$, $L = 8 \cdot 10^{-5} m$, $\rho_1 = 1000\, kg/m^3$, $\rho_2 = 1 \,kg/m^3$, $\mu_1 = 10^{-3}\, Ns/m^2$ and $\mu_2 = 10^{-5}\, Ns/m^2$. For the Cahn-Hilliard equation we used $\varepsilon = 4h$ with $h$ being the initial grid width. Figure \ref{fig:droplet} shows the droplet simulation at different stages. $144$ cores were used for one simulation and the average run time per simulation was $3$ hours. In total, $16$ simulations were carried out with volume ratios of the two droplets varying between $0\%$, $3.\bar{3}\%$, $6.\bar{6}\%$, and $10\%$. The same initial data variation was applied for the initial droplet velocities.

\subsection*{\name algorithm}
\label{subsec:algorithm}
As shown in \refFig{main}(b), \name comprises three networks.
The first network is an encoder (\encoder), implemented using \acp{CNN}. 
The encoder sees all of the input projection images and extracts latent features of the sample from the projections.
By transferring knowledge across different timestamps, it learns the general features of the sample, which is crucial for accurate 3D reconstruction from sparse projections.
The second network is an IoR generator (\generator), formed by a \ac{MLP}.
It learns the mapping from 4D spatial-temporal coordinates
(\spatialCoordinates, \temporalCoordinates) to the refractive index of the sample \refractiveIndex($\delta$, $\beta$) , with the assistance of the encoder.
Here, $\delta$ and $\beta$ are real and positive numbers representing the real and imaginary parts of the complex refractive index, respectively.
The generator takes as input the extracted latent features from the encoder and the 4D coordinates at each spatial-temporal point.
It is trained to output the value of the refractive index at each 4D point.
Predicted projections can be calculated by integrating the output refractive index along the line of propagation, following the law of X-ray propagation and interaction with matter~\cite{Paganin2006CoherentX-ray}.
\change{}{
\refFig{render_new_prediction} illustrates the process of novel view prediction.}
The third network is a \ac{CNN} discriminator (\discriminator).
The discriminator learns to minimize the difference between the image patches from the real projections and the predictions generated by \name.
It sees both the image patches from the real projections and the \name predictions, and it learns to distinguish the fake ones from the real images.
The encoder and the generator are optimized based on the feedback from the discriminator.
They are trained to fool the discriminator by generating indistinguishable images, leading to high-quality 3D reconstruction of the sample.

\myfigure{render_new_prediction}{\protect\change{}{
Novel view prediction from \name.
The IoR Generator calculates the refractive index at each 4D point (\spatialCoordinates, \temporalCoordinates).
For each time point, multiple query points are taken along a specific ray direction to obtain the refractive index \refractiveIndex($\delta$,$\beta$) along that ray.
The refractive index is integrated along the ray using the physical principles of X-ray interaction with matter.
This integration provides the value of each pixel in the predicted image.
By utilizing more rays, we can generate a projection image.
}}

\name is trained by optimizing a loss function based on adversarial learning, as expressed in \refEq{adv_loss}.
\begin{equation}
{\mathcal L_\mathrm{GAN}} = \mathbb{E}_{\realContrast \sim \dataDistribution}\log(\discriminator(\realContrast))+\mathbb{E}_{\predictContrast\sim \randomDistribution}\log(1-\discriminator (\predictContrast))~,
\label{eq:adv_loss}
\end{equation}
where \realContrast and \predictContrast refer to image patches from the real and predicted projections, respectively. 
The simulation dataset included both absorption and phase contrast, resulting in two-channel representations of \realContrast and \predictContrast.
Since the phase contrast in our simulations was directly proportional to the absorption contrast, only the absorption images were shown in the present work.
In the case of the experimental data, only absorption-contrast images were obtained. Therefore, \realContrast and \predictContrast are representations of absorption contrast.
The discriminator is trained to minimize the difference between these two patches.
The data distribution, denoted as \dataDistribution, represents the distribution over the collected projections in our experiments, which are considered real projections by the discriminator. 
On the other hand, \randomDistribution represents the distribution over all generated predictions, which are considered fake projections by the discriminator. 
$ \mathbb E$ represents the expectation of a function. 
By minimizing the difference between the real and predicted projections, the discriminator provides feedback to the generator and the encoder networks, enabling them to generate more accurate and realistic 3D reconstructions.

\subsection*{Network and training details}
\label{subsec:training details}
For the implementation, we used ResNet34~\cite{he2016deep} as encoders and PatchGAN discriminator~\cite{ledig2017photo} as the discriminator. 
The generator was built by five layers of ResBlocks~\cite{he2016deep}, where the first three layers contain multiple parallel weight-sharing ResBlocks, with the number of parallel blocks equals to the number of constraints used.
An average pooling operation was applied after the first three layers to take the average, and the last two layers introduced new learning parameters to the networks.
We used the same sampling method as in \cite{schwarz2020graf} to extract image patches, where each image patch was formed by sampling a $32\times32$ square grid with a flexible scale, position, and stride.

\ac{GAN}s are known to have the problem of local equilibria and mode collapse~\cite{goodfellow2016deep, kodali2017convergence}.
Therefore, in the first five epochs of the training, we force \name to learn only the self-consistency over the two recorded projections.
This consistency is enforced by optimizing an \ac{MSE} loss function between the recorded projections and the network predictions, as expressed in \refEq{mse_loss}.
\begin{equation}
\mathcal L_\mathrm{MSE} = \sum_{\change{}{\viewAngle = \realAngle}} \left \|\realContrast - \predictContrast \right \|^2_2,
\label{eq:mse_loss}
\end{equation}
where \realContrast and \predictContrast denote image patches from the real and predicted projections, 
\change{}{while \realAngle and \viewAngle stand for the view angle of the recorded projections and the predictions, respectively.}
The adversarial loss was applied starting from the sixth epoch, once a convergence was obtained from the consistency of the \change{}{constraining} projections.
The ADAM optimizer ~\cite{kingma2014adam} with a mini-batch size of 8 was used throughout the training. 
\change{}{For the training of water droplet datasets, }
We set the learning rates to be 0.0001 for the networks and decayed the learning rate by 0.1 every 100 epochs. 
The results presented in the \nameref{subsec:simulation_results} section were the results of 200 epochs, which took around 70 hours of training on a NVIDIA V100 GPU with 32 GB of RAM.
\change{}{
For the experimental additive manufacturing data, we employed a random dice mode for training, where half of the iterations used GAN loss and the other half used MSE loss.
The learning rate was set to 0.0001, and no decay was applied. 
The results presented in the \nameref{subsec:exp result} section were based on 400 epochs of training, which took 35 hours on an NVIDIA A100 GPU with 80 GB of RAM.}

\section*{Data availability}
The data that support the findings of this study is available at \href{https://figshare.com/s/bca51069fd2e287ef3c9}{figshare}.

\section*{Code availability}
The code developed in this study is available at \href{https://github.com/yuhez/4D-ONIX}{https://github.com/yuhez/4D-ONIX}.

\section*{Acknowledgement}
We are grateful to Z. Matej for his support and access to the GPU‐computing cluster at MAX IV. 
\change{}{We are also grateful to Malgorzata Makowska, et al., for sharing the additive manufacturing data~\cite{aluminadata2023} to demonstrate the capabilities of 4D-XMPI. 
This data was acquired within a frame of the project ``Operando tomography of Selective Laser Additive Manufacturing" funded from the SNF Spark grant CRSK-2\_196085.}
This work has received funding by ERC‐2020‐STG 3DX‐FLASH 948426 and the HorizonEIC-2021-PathfinderOpen-01-01, MHz-TOMOSCOPY 101046448.

\bibliographystyle{unsrt}
\bibliography{main}

\end{document}


\maketitle

\begin{abstract}
This document provides supplementary information for ``\name for reconstructing 3D movies from sparse X-ray projections via deep learning.” 
In this material, we report supplementary figures and descriptions to complement the main article.
\end{abstract}

\section{Reconstruction for reproducible processes with variable number of experiments for simulated water droplet collision}

In this section, we investigate the performance of \name by training it with multiple experiments of reproducible processes.
Specifically, we randomly generated 16 projection pairs in a \ang{180} range from a simulated droplet collision dataset.
The dataset comprises 75 timestamps, which capture the centered collision between two water droplets.
The angle between the two projections within each pair is \ang{23.8}.
Here, we refer to each unique projection pair as an ``experiment".

We trained \name using 1, 2, 4, 8, and 16 experiments separately. 
For the 16 experiments, the $\varphi_1$ angles (azimuthal angles of the first projections, see Figure 2 of the main article) were: \ang{0}, \ang{2}, \ang{13}, \ang{16}, \ang{26}, \ang{28}, \ang{43}, \ang{52}, \ang{64}, \ang{74}, \ang{87}, \ang{95}, \ang{102}, \ang{115}, \ang{130}, \ang{144}.
Every second experiment was used for the training with eight experiments.
For the training with four experiments, every fourth experiment was employed.
The first and the ninth experiments were used for the training with two experiments. 
All experiments were used for the training with 16 experiments, while only the first experiment was used for the single-experiment training. 
The precise angles used in the training are detailed in Table~\ref{tbl:single_exp_results}.
Please note that these angles are unknown to \name during the training process.
The results were obtained after 800, 800, 800, 400, and 200 epochs for the training with 1, 2, 4, 8, and 16 experiments, respectively,
Throughout all training sessions, we set an initial learning rate of 0.0001 and decayed the learning rate by 0.1 in the middle of the total epochs for each training.
The rest of the model parameters are the same as those used in the main paper. 

\begin{table}[htbp!]
\centering
\caption{Comparison of \name reconstructions trained with different numbers of experiments for reproducible processes. The yellow, green, blue, and brown lines represent the additional projections introduced as the number of experiments increases.
}
  \label{tbl:single_exp_results}
 \vspace{.2cm}%
\includegraphics[width=0.99 \linewidth]{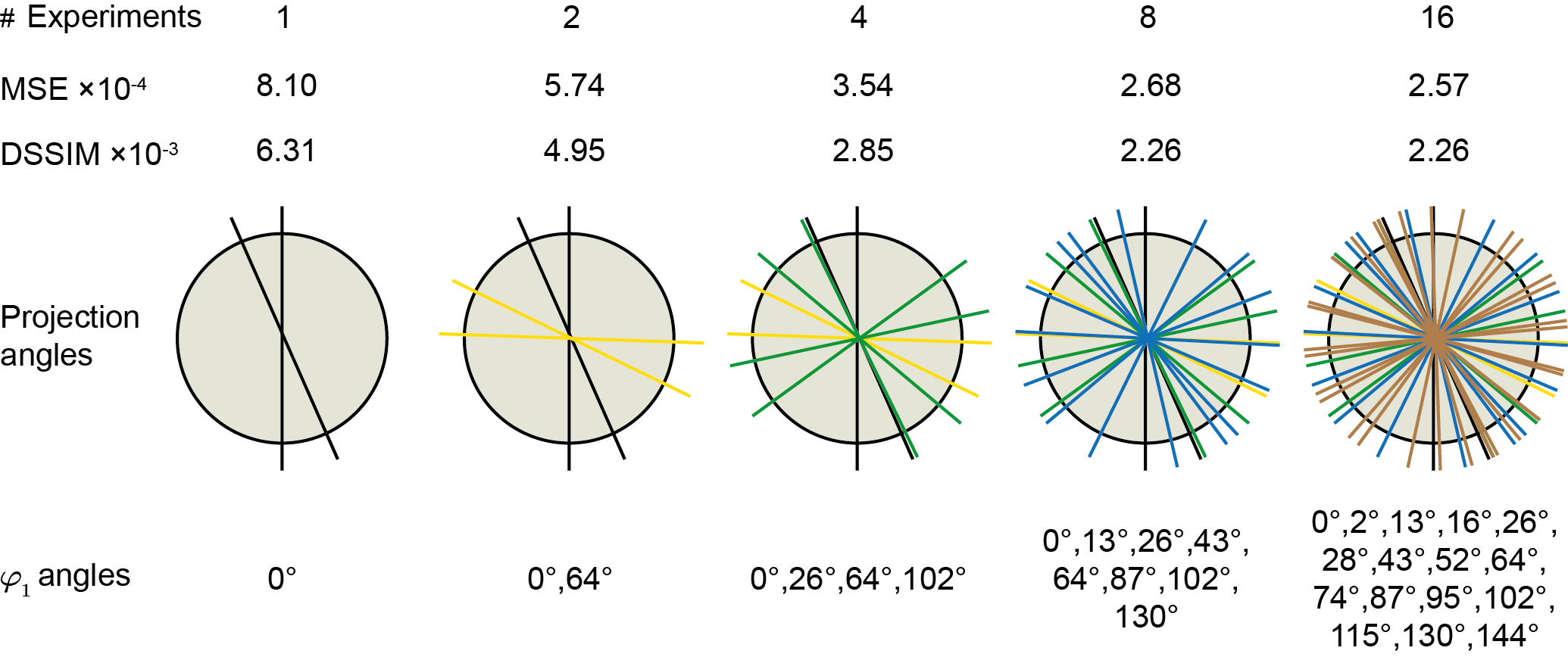}
\end{table}

\begin{figure}
\centering
\includegraphics[width=1.0 \linewidth]{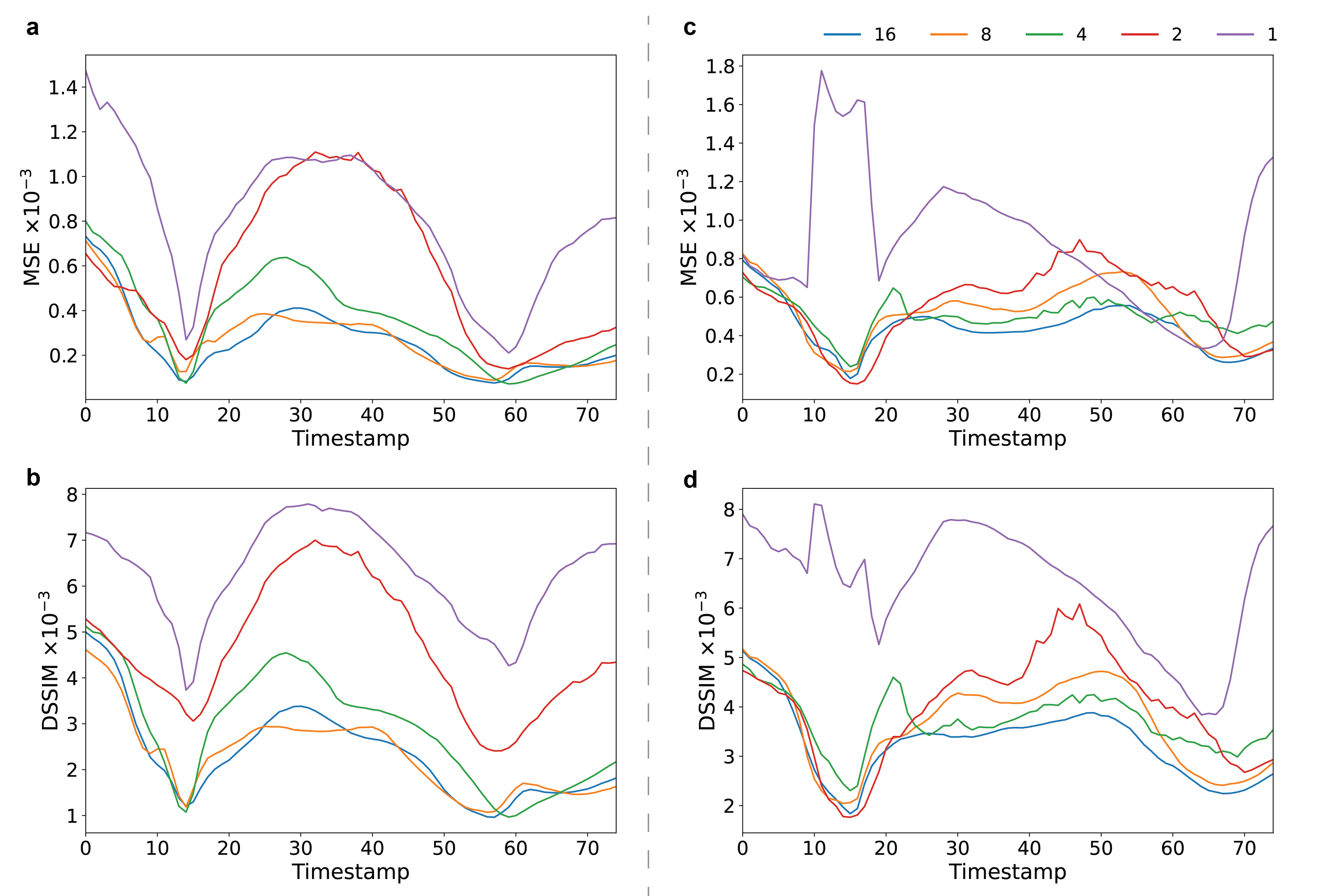}
\caption{Comparison of error distribution with time for \name reconstructions trained with 16 (blue), 8 (orange), 4 (green), 2 (red), and 1 (purple) experiments for a reproducible process (\textbf{a, b}) and a quasi-reproducible process (\textbf{c,d}).
\textbf{a} Mean Squared Error (MSE)  and \textbf{b} Dissimilarity Structure Similarity Index Metric (DSSIM) distribution with time for \name reconstructions trained with different numbers of experiments for a reproducible process. \textbf{c} MSE and \textbf{d} DSSIM distribution with time for \name reconstructions trained with different numbers of experiments for a quasi-reproducible process. 
}
\label{fig:error_combine_x_frame}
\end{figure}

\begin{figure}[htbp!]
\centering
\includegraphics[width=1 \linewidth]{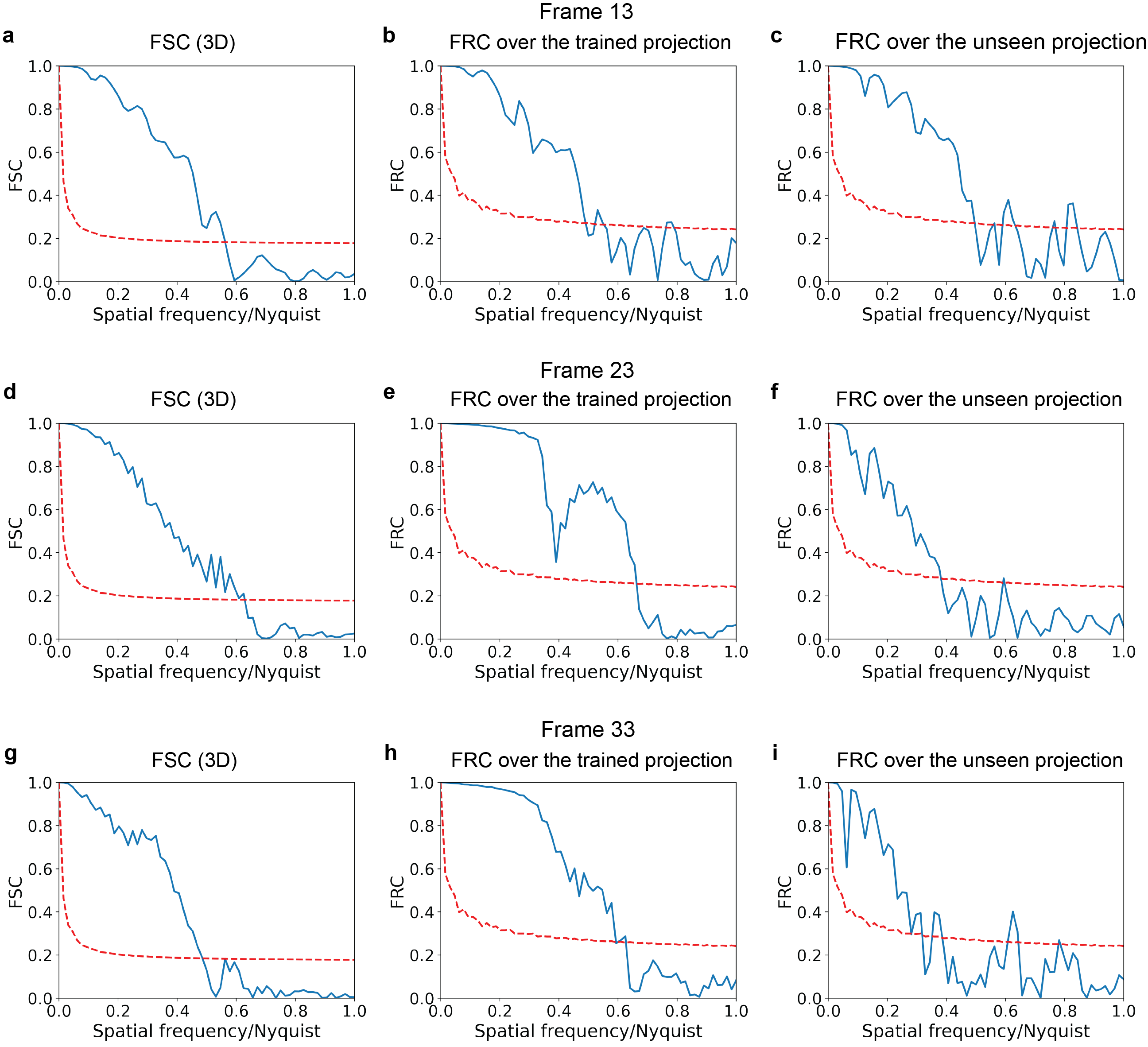}
\caption{Demonstration of the retrieved resolution for the \name reconstructions trained with 16 experiments of reproducible processes.
The blue curves represent the 3D Fourier Shell Correlation (FSC) and 2D Fourier Ring Correlation (FRC) across different spatial frequencies, while the red dashed curves indicate the half-bit criterion.
Three example timestamps are shown.
\textbf{a} shows the \ac{FSC} between the \name reconstruction and the 3D ground truth for timestamp 13, which gives a resolution of around 3.5 voxels. \textbf{b} and \textbf{c} show for the same timestamp the 2D \ac{FRC} for the seen projection and the unseen projection, respectively, which correspond to a resolution of 4 pixels for both. 
\textbf{d} the \ac{FSC} for timestamp 23, which gives a resolution of around 3 voxels. \textbf{e} and \textbf{e} the 2D \ac{FRC} for the seen projection and the unseen projection for timestamp 23, which corresponds to a resolution of 3 pixels and 5 pixels, respectively. 
\textbf{g} the \ac{FSC} for timestamp 33, which gives a resolution of 4 voxels. \textbf{h} and \textbf{i} the 2D \ac{FRC} for the seen projection and the unseen projection for timestamp 33, which corresponds to a resolution of 3 pixels and 7 pixels, respectively. 
}
\label{fig:FSC}
\end{figure}

\begin{figure}[htbp!]
\centering
\includegraphics[width= \linewidth]{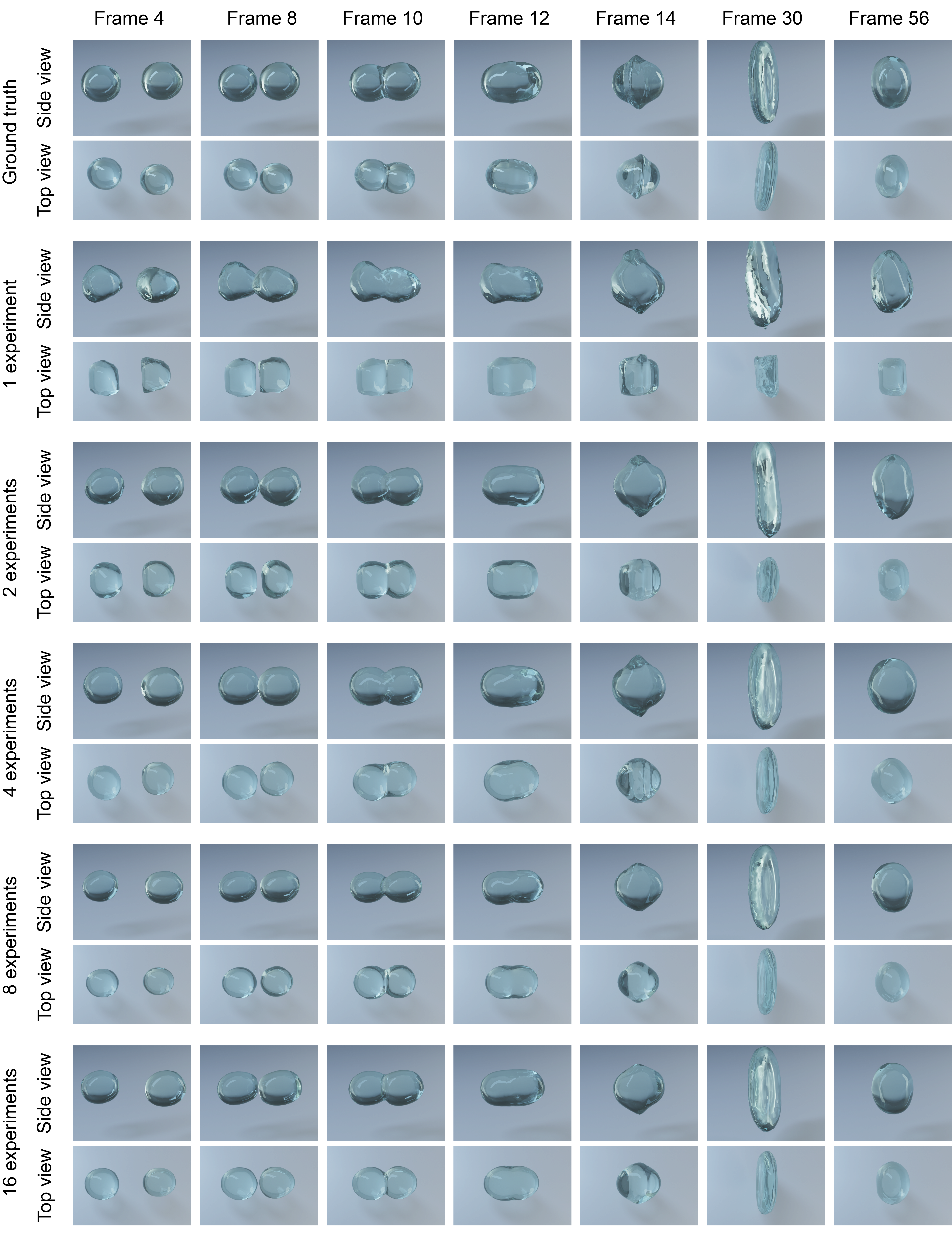}
\caption{Demonstration of \name reconstructions trained with different numbers of experiments for reproducible processes. 
For the selected timestamps, the 3D rendering from the side and top views of the ground truth and the \name reconstructions using different numbers of experiments are presented.
}
\label{fig:simulation_single_exp_compare_exp}
\end{figure}

We evaluated the performance of the 4D reconstructions reconstructed using different numbers of experiments.
The performance evaluation calculates the difference between the \name output and the ground truth. 
Please note that the ground truth is never shown to the \name model at any stage, and it is used only for evaluation.
First, we calculated the 4D \ac{MSE} and the \ac{DSSIM}~\cite{Wang2004SSIM} between the ground truth and the \name reconstructions, as shown in Table~\ref{tbl:single_exp_results}.
Next, we calculated the distribution of 3D \ac{MSE} and \ac{DSSIM} with time, as shown in Figure~\ref{fig:error_combine_x_frame}\textbf{a, b}.
The spatial resolution of the reconstructions with 16 experiments was evaluated using \ac{FSC} and \ac{FRC}, which gave the resolution in 3D and 2D, respectively~\cite{saxton1982correlation,VANHEEL2005250}.
The results of three example timestamps at different stages of the collision are shown in Figure~\ref{fig:FSC}.
The \ac{FRC} was calculated over the trained projection and the unseen projection (perpendicular to the beam plane) separately. 
The reconstruction results trained with 1, 2, 4, 8, and 16 experiments, along with the ground truth, are presented in Figure~\ref{fig:simulation_single_exp_compare_exp}, Supplementary Movie 1, and Supplementary Movie 2.

\section{Reconstruction for quasi-reproducible processes with a variable number of experiments for simulated water droplet collision}
In this section, we assessed the performance of \name by training it with multiple experiments involving quasi-reproducible processes. 
We conducted 16 simulated droplet collision experiments, each comprising 75 timestamps depicting the collision of two droplets with a 10\% variation in droplet size and collision velocity.
From each simulation experiment, we generated a projection pair. 
The angles of the projections were the same as the ones used for the reproducible processes, and the angle between the two projections remains consistent at \ang{23.8}.

\begin{figure}[htbp!]
\centering
\includegraphics[width=1 \linewidth]{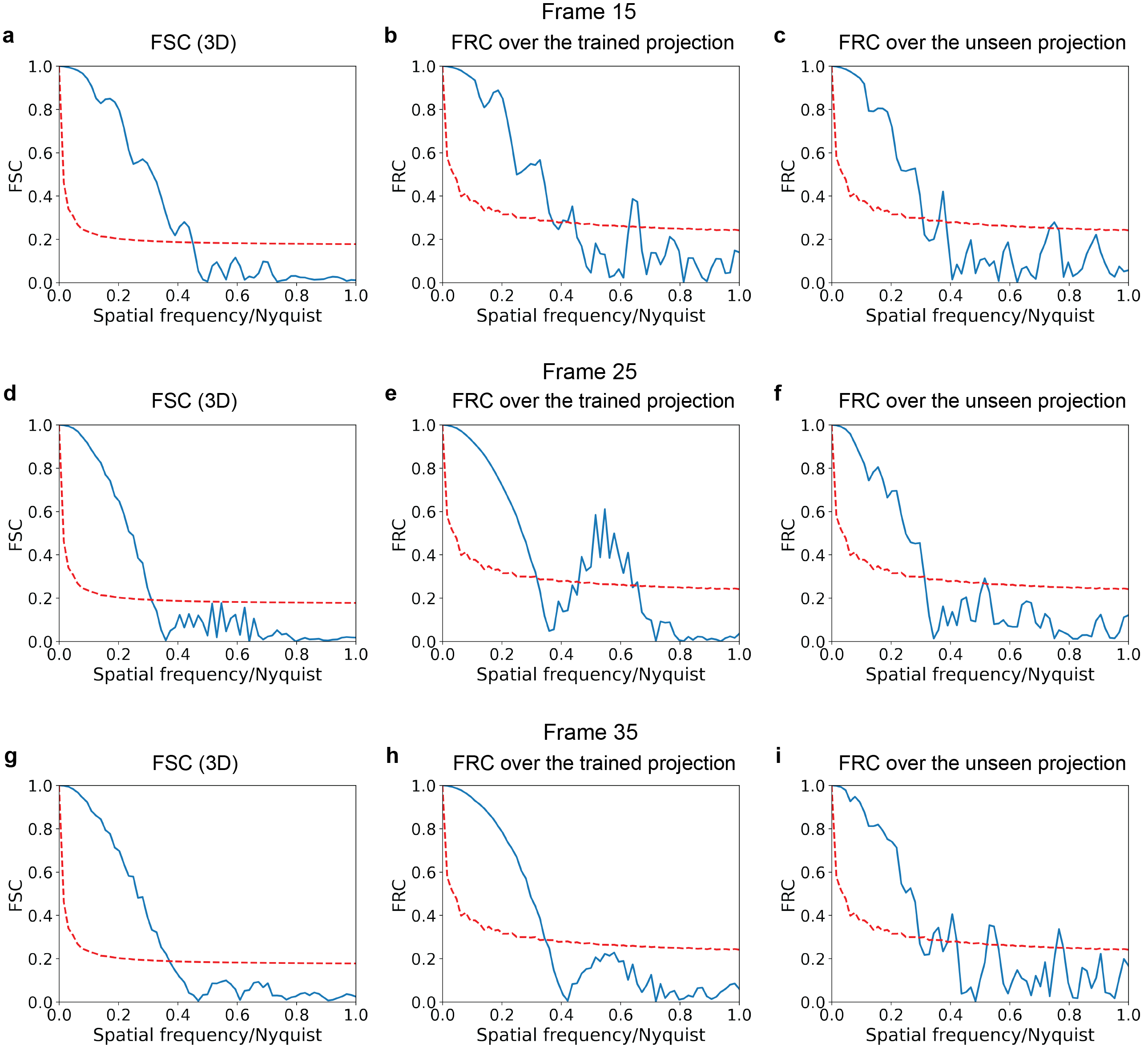}
\caption{Demonstration of the retrieved resolution for the \name reconstructions trained with 16 experiments of quasi-reproducible processes.
The blue curves represent the 3D Fourier Shell Correlation (FSC) and 2D Fourier Ring Correlation (FRC) across different spatial frequencies, while the red dashed curves indicate the half-bit criterion.
Three example timestamps are shown.
\textbf{a} shows the \ac{FSC} between the \name reconstruction and the 3D ground truth for timestamp 15, which gives a resolution of around 4 voxels. \textbf{b} and \textbf{c} show for the same timestamp the 2D \ac{FRC} for the seen projection and the unseen projection, respectively, which correspond to a resolution of 5 pixels and 6 pixels, respectively.
\textbf{d} the \ac{FSC} for timestamp 25, which gives a resolution of around 6 voxels. \textbf{e} and \textbf{e} the 2D \ac{FRC} for the seen projection and the unseen projection for timestamp 25, which corresponds to a resolution of 6 pixels for both. 
\textbf{g} the \ac{FSC} for timestamp 35, which gives a resolution of 5 voxels. \textbf{h} and \textbf{i} the 2D \ac{FRC} for the seen projection and the unseen projection for timestamp 35, which corresponds to a resolution of 6 pixels and 7 pixels, respectively. 
}
\label{fig:FSC_multi}
\end{figure}

\begin{figure}[htbp!]
\centering
\includegraphics[width= \linewidth]{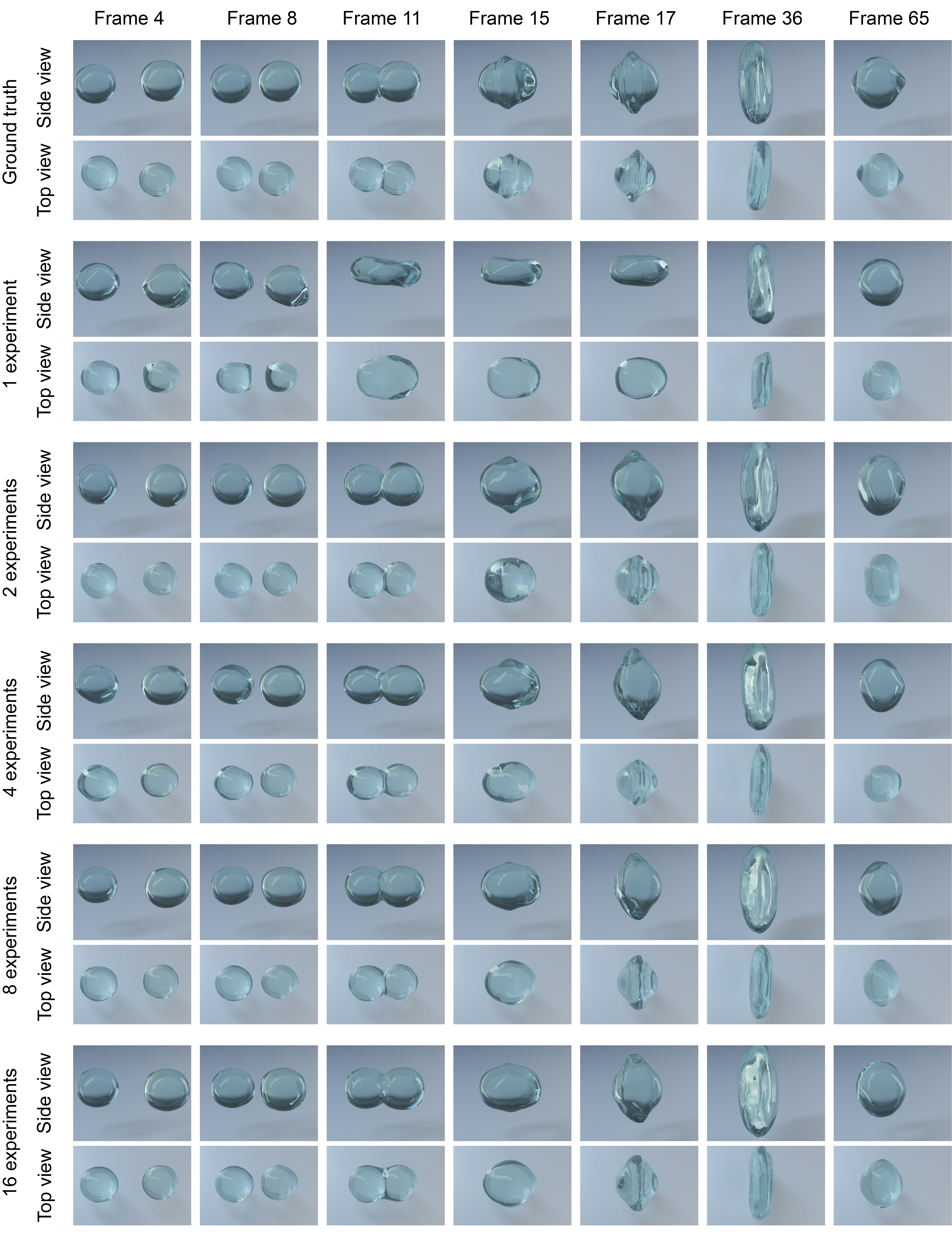}
\caption{Demonstration of \name reconstructions trained with different numbers of experiments for quasi-reproducible processes. 
For the selected timestamps, the 3D rendering from the side and top views of the ground truth and the \name reconstructions using different numbers of experiments are presented.
}
\label{fig:simulation_multiple_exp_compare_exp}
\end{figure}

We trained \name separately using 1, 2, 4, 8, and 16 experiments and assessed the performance of the 4D reconstructions obtained with different numbers of experiments.
The training parameters are the same as reported in the previous section.
The 4D \ac{MSE} and the \ac{DSSIM} between the ground truth and the \name reconstructions are shown in Table~\ref{tbl:multiple_exp_results}.
Next, we calculate the distribution of 3D \ac{MSE} and \ac{DSSIM} with time, as shown in Figure~\ref{fig:error_combine_x_frame} \textbf{c, d}.
Examples of the \ac{FSC} and \ac{FRC} curves for three timestamps at different stages of the collision are shown in Figure~\ref{fig:FSC_multi}.
Same as before, the \ac{FRC} was calculated over the trained projection and the unseen projection (perpendicular to the beam plane) separately. 
Figure~\ref{fig:simulation_multiple_exp_compare_exp}, Supplementary Movie 3, and Supplementary Movie 4 show the ground truth and the reconstruction results trained with 1, 2, 4, 8, and 16 experiments.

\begin{table}[htbp!]
\centering
\caption{Comparison of \name reconstructions trained with different numbers of experiments for quasi-reproducible processes.
\label{tbl:multiple_exp_results}}
\setlength{\tabcolsep}{4pt}
\begin{tabular}{cccc}
\hline
\# Experiments  & $\varphi_1$ angles & \begin{tabular}[c]{@{}c@{}}MSE\\ $\times10^{-4}$\end{tabular}  & \begin{tabular}[c]{@{}c@{}}DSSIM\\ $\times10^{-3}$\end{tabular} \\
\hline
1       &   \ang{0}        & 8.64              & 6.46 \\ 
2      &   \ang{0}, \ang{64}        & 5.45              & 4.05 \\
4       &    \ang{0}, \ang{26},\ang{102},\ang{64}        & 4.84              & 3.67 \\
8       &    \ang{0}, \ang{13},\ang{26}, \ang{43}, \ang{64}, \ang{87},\ang{102}, \ang{130}        & 4.98              & 3.66 \\
16     &  \begin{tabular}[c]{@{}c@{}}\ang{0}, \ang{2}, \ang{13}, \ang{16}, \ang{26}, \ang{28}, \ang{43}, \ang{52},\\ \ang{64}, \ang{74}, \ang{87}, \ang{95}, \ang{102}, \ang{115}, \ang{130}, \ang{144}\end{tabular}        & 4.29              & 3.24 \\
\hline
\end{tabular}
\end{table}

\newpage

\section{Data preparation and results of \name on experimental additive manufacturing data}
As reported in the main article, we mimic multiple XMPI experiments using a time-resolved tomographic dataset of the remelting process of the magnetite-modified alumina.
The original time-resolved tomographic dataset has the shape of $60\times200\times960\times180$, where 60 is the number of time stamps, 200 is the number of projections, and  $960\times180$ is the dimension of the measured projection. 
We first applied flat-field correction to remove the flat-field noise of the projections, and next applied a Paganin filter to enhance the contrast and mitigate the noise on the measured intensity images~\cite{paganin2002simultaneous}.
Then, we selected a $960\times64$ region where the remelting dynamics happen and resized the selected area to $128\times64$ for faster computation.

For each tomogram, a 3D ``ground truth" was prepared using the gridrec tomographic reconstruction algorithm~\cite{dowd1999developments} on the full set of 200 projections. 
This ground truth serves as a reference to validate \name's performance.

\begin{figure}[htbp!]
\centering
\includegraphics[width= 0.7\linewidth]{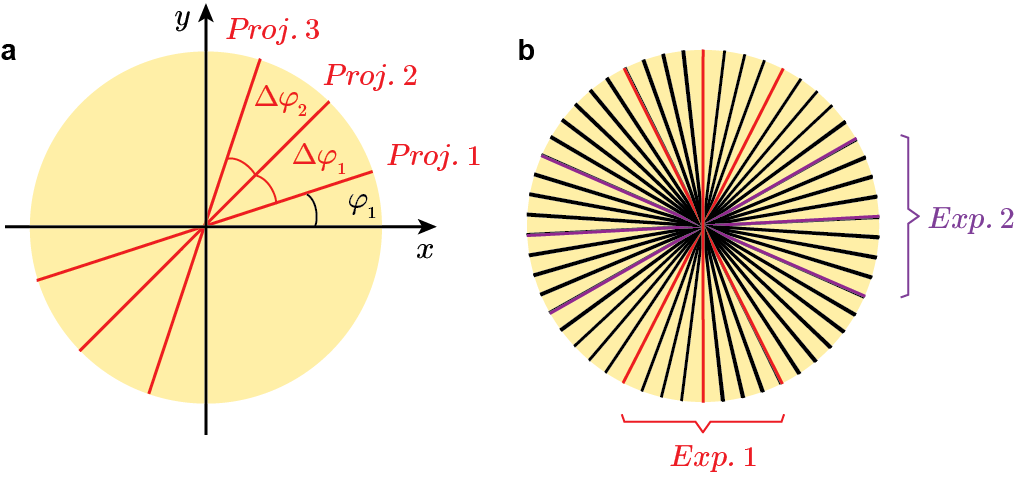}
\caption{
\textbf{a} Geometry of the simulated \acf{XMPI} experiment. 
Three projections (red lines) were selected from the tomographic dataset to replicate an XMPI measurement, with equal angular intervals of $\Delta \varphi_1 = \Delta \varphi_2 = \ang{27}$.
\textbf{b} Multiple XMPI experiments can be simulated using the 200 projections from the tomographic dataset.
In this example, we demonstrate two experiments (colored red and purple), each consisting of three projections, separated by \ang{27}. 
These experiments are treated as independent, and thus, the relative angles between them are not used as prior information.
}
\label{fig:exp_angles}
\end{figure}

\begin{figure}[htbp!]
\centering
\includegraphics[width= 0.6\linewidth]{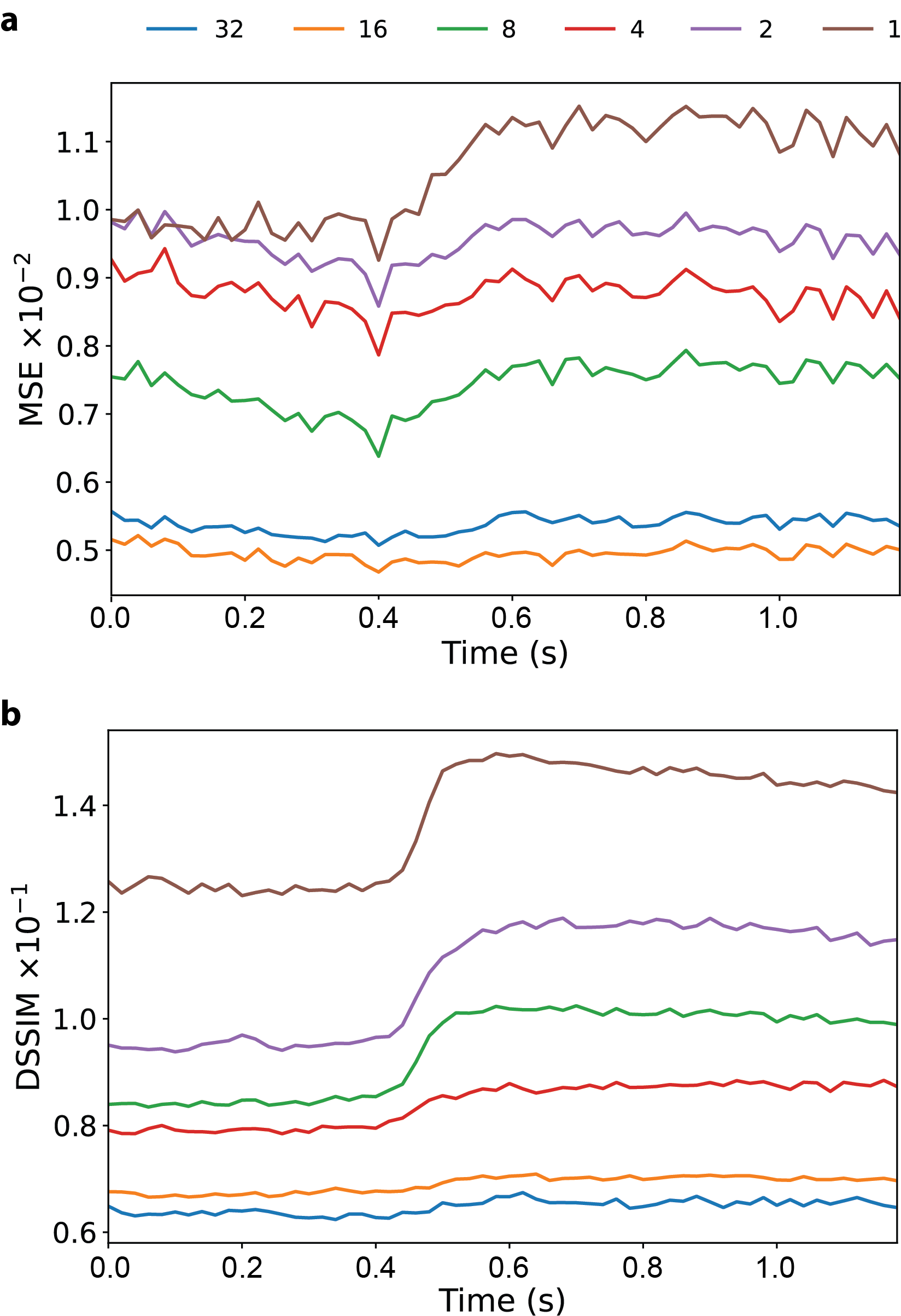}
\caption{Comparison of \textbf{a} Mean Squared Error (MSE)  and \textbf{b} Dissimilarity Structure Similarity Index Metric (DSSIM) distribution with time for \name reconstructions trained with different numbers of experiments for the experimental additive manufacturing data.
}
\label{fig:error_time}
\end{figure}

To train \name, we simulated multiple XMPI experiments by extracting ultra-sparse projections from the tomographic dataset. 
As shown in Figure \ref{fig:exp_angles}, we selected three projections per experiment, where the geometry is similar to the XMPI setup reported in \cite{asimakopoulou2023development}.
The angle of the first projection is denoted as $\varphi_1$, consistent with the notation used in Figure 2\textbf{a} of the main article. 
The relative angles between the first and second projections and between the second and third projections are represented by $\Delta \varphi_1$ and $\Delta \varphi_2$, respectively.
In this case, we use $\Delta \varphi_1 = \Delta \varphi_2 = \ang{27}$.
For instance, if $\varphi_1 = 0$, the three projection angles are \ang{0}, \ang{27}, and \ang{54}.
We evaluated the performance of \name trained with different numbers of experiments, ranging from 1, 2, 4, 8, 16, to 32. 
The $\varphi_1$ angles of the selected experiments are reported in Table~\ref{tbl:additive_manufacturing}, along with the 4D MSE and DSSIM values comparing the ground truth and the \name reconstructions.
The evolution of 3D MSE and DSSIM values over time is presented in Figure \ref{fig:error_time}.

\begin{table}[htbp!]
\centering
\caption{Comparison of \name reconstructions trained with different numbers of experiments for experimental additive manufacturing data.
\label{tbl:additive_manufacturing}}
\setlength{\tabcolsep}{4pt}
\begin{tabular}{|c|c|c|c|}
\hline
\# Experiments  & $\varphi_1$ angles & \begin{tabular}[c]{@{}c@{}}MSE\\ $\times10^{-3}$\end{tabular}  & \begin{tabular}[c]{@{}c@{}}DSSIM\\ $\times10^{-2}$\end{tabular} \\
\hline
1       &   \ang{0}        & 10.6              & 14.2 \\ 
\hline
2      &   \ang{0}, \ang{64.8}        & 10.2              & 9.7 \\
\hline
4       &    \ang{0}, \ang{32.4},\ang{64.8},\ang{97.2}        & 8.8              & 8.4 \\
\hline
8       &    \ang{0}, \ang{16.2},\ang{32.4}, \ang{48.6}, \ang{64.8}, \ang{81},\ang{97.2}, \ang{113.4}        & 7.4              & 9.5 \\
\hline
16     &  \begin{tabular}[c]{@{}c@{}}\ang{0}, \ang{8.1}, \ang{16.2}, \ang{24.3}, \ang{32.4}, \ang{40.5}, \ang{48.6}, \ang{56.7},\\ \ang{64.8}, \ang{72.9}, \ang{81}, \ang{89.1}, \ang{97.2}, \ang{105.3}, \ang{113.4}, \ang{121.5}\end{tabular}        & 5.1              & 6.9 \\
\hline
32     &  \begin{tabular}[c]{@{}c@{}}\ang{0}, \ang{3.6}, \ang{7.2}, \ang{10.8}, \ang{14.4}, \ang{18}, \ang{21.6}, \ang{25.2},\\ \ang{28.8}, \ang{32.4}, \ang{36}, \ang{39.6}, \ang{43.2}, \ang{46.8}, \ang{50.4}, \ang{54}\\ \ang{71.1},  \ang{74.7},  \ang{78.3},  \ang{81.9},  \ang{85.5},  \ang{89.1},  \ang{92.7},  \ang{96.3}, \\ \ang{99.9},
       \ang{103.5}, \ang{107.1}, \ang{110.7}, \ang{114.3}, \ang{117.9}, \ang{121.5}, \ang{125.1}
\end{tabular}        & 5.2             & 6.9 \\

\hline
\end{tabular}
\end{table}

\section{Comparison with state-of-the-art methods}

In this section, we evaluate the performance of \name by comparing it with two other reconstruction approaches: (i) simultaneous algebraic reconstruction technique (SART)~\cite{andersen1984simultaneous}, a classic iterative approach for sparse-view reconstructions, and (ii) \ntoi~\cite{hendriksen2020noise2inverse}, a state-of-the-art deep-learning approach for sparse-view tomographic reconstruction. 

\begin{figure}[htbp!]
\centering
\includegraphics[width= 0.8\linewidth]{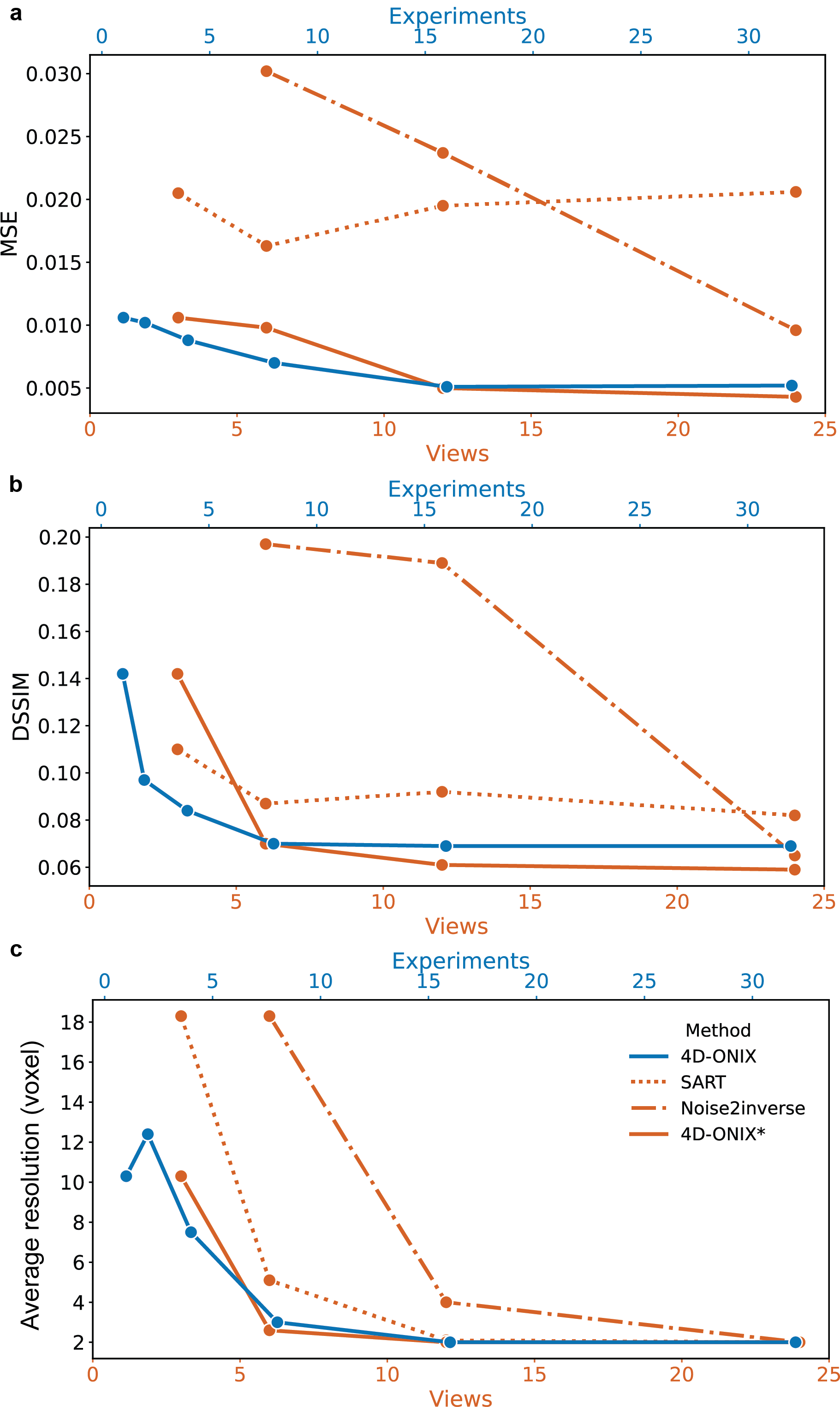}
\caption{ Comparison of \textbf{a} \acf{MSE}, \textbf{b} \acf{DSSIM}, and \textbf{c} average resolution determined by \acf{FSC} for the reconstructions of experimental additive manufacturing data using different methods.
The bottom x-axis (orange) and orange lines represent the results of \acf{SART}, \ntoi, and \namestar, reconstructed with varying numbers of projections. The upper x-axis (blue) and blue lines represent the results of \name, reconstructed by combining different numbers of experiments, each using three projections.
Note that the first data points of \name and \namestar represent the same experiment, as \namestar trained with three projections corresponds to \name trained with a single experiment.
}
\label{fig:compare_sart_n2i}
\end{figure}

\begin{figure}[htbp!]
\centering
\includegraphics[height= 0.76\paperheight]{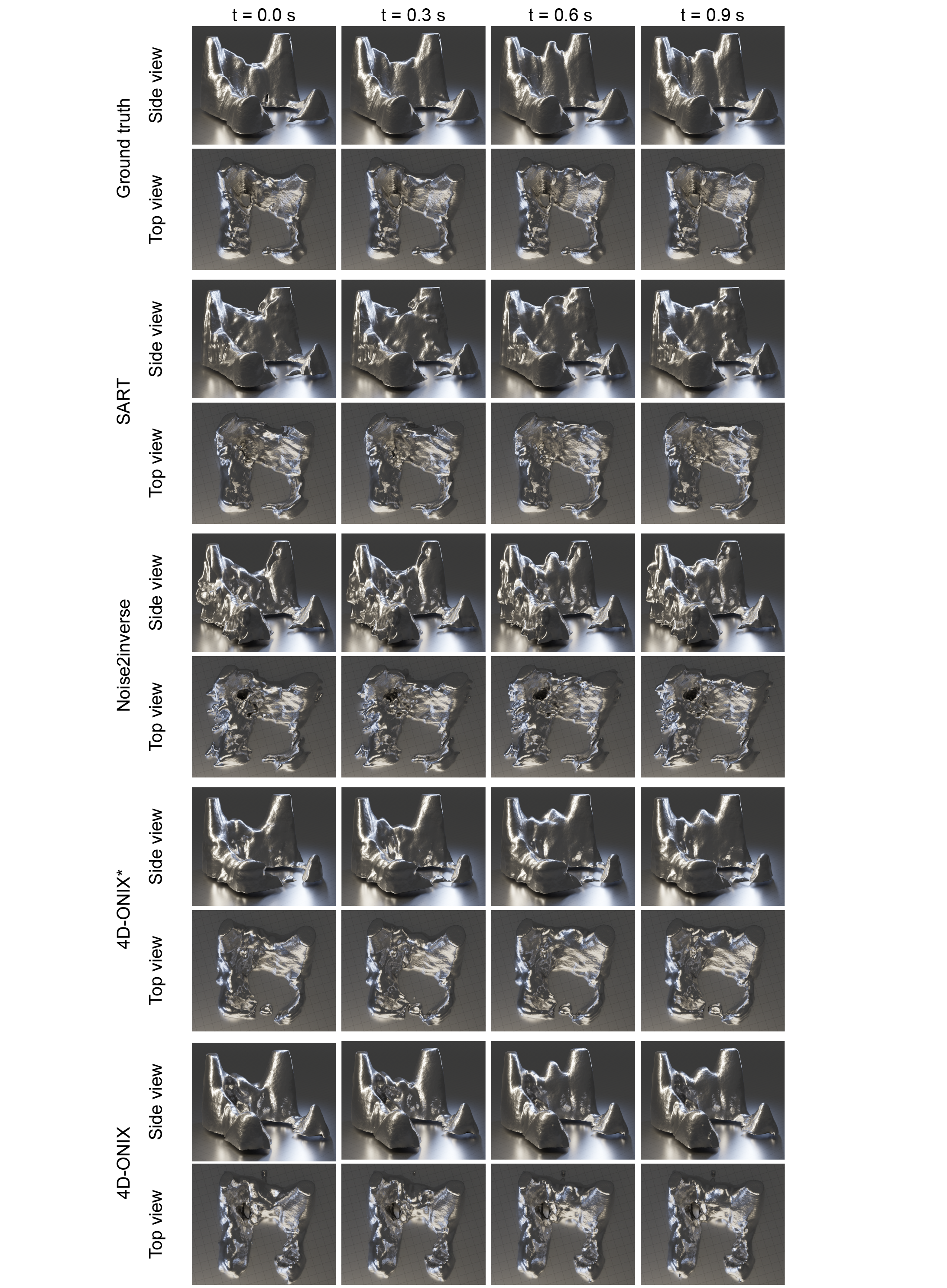}
\caption{ Demonstration of \name reconstructions comparing with \acf{SART}, \ntoi, \namestar, and ground truth.
For the selected time points, the 3D rendering of the reconstructed results from the side and top views are presented.
The results of SART, \ntoi, and \namestar are reconstructed using 24 projections.
The results of \name are reconstructed by combining 32 experiments of 3 projections.
}
\label{fig:additive_manufacturing_compare_results}
\end{figure}

We use the experimental additive manufacturing data for the evaluation. 
Since these methods lack the capacity to generalize across different experiments, we evaluate using a hypothetical scenario where more projections were used to reconstruct 4D from a single experiment.
To have a fair comparison, we also show the results of \name trained by using more projections of a single experiment, as denoted by \namestar.
The results are shown in Figure \ref{fig:compare_sart_n2i} and Figure \ref{fig:additive_manufacturing_compare_results}. 

In Figure~\ref{fig:compare_sart_n2i}, the orange lines and the bottom x-axis show the results of SART, \ntoi, and \namestar, reconstructed using 3, 6, 12, and 24 projections for a single 4D experiment. The blue lines and the upper x-axis represent the results of \name, reconstructed by combining different numbers of experiments, each with three projections, as summarized in Table \ref{tbl:additive_manufacturing}.
Figure \ref{fig:additive_manufacturing_compare_results} shows 3D renderings of the reconstructions from different methods alongside the ground truth. 
To achieve optimal visualization, a specific threshold was applied to the rendering of each method.
As illustrated in Figures \ref{fig:compare_sart_n2i} and \ref{fig:additive_manufacturing_compare_results}, \name trained with 32 experiments achieves comparable performance to SART, \ntoi, and \namestar trained on single experiments with 24 projections. 
However, all reconstructions exhibit artifacts due to the noisy nature of the experimental projections.

\bibliographystyle{unsrt}
\bibliography{Biblio}